\documentclass[USenglish,twocolumn]{article}

\usepackage[sort&compress,square,numbers]{natbib}
\usepackage[big,online]{template}

\addto{\captionsenglish}{%
  \renewcommand{\refname}{References}
}

\usepackage{ragged2e}
\usepackage{gensymb}
\theoremstyle{dgthm}

\usepackage{lipsum}
\usepackage{multirow}
\usepackage{amsmath}
\usepackage{float}
\usepackage{stfloats} 
\usepackage{mathtools}
\usepackage[footnotehyper]{nicematrix}

\newcommand{\upperRomannumeral}[1]{\uppercase\expandafter{\romannumeral#1}}

\begin{document}

\articletype{Research Article}

\justifying

\author*[1]{Xin Mu}
\author[2]{Frank Weiss}
\author[4]{Hongyao Chua}
\author[5]{Robert Lawrowski}
\author[2]{Jared C. Mikkelsen}
\author[2]{John N. Straguzzi}
\author[2]{Hannes Wahn}
\author[2]{Piyush Kumar}
\author[4]{Guo-Qiang Lo}
\author[3]{Joyce K. S. Poon}
\author[5]{Mariel Jama}
\author*[1]{Wesley D. Sacher}

\affil[1]{
\newline\textbf{Xin Mu}, Max Planck Institute of Microstructure Physics, Weinberg 2, 06120 Halle, Germany; and University of Toronto, 10 King’s College Road, Toronto, Ontario, M5S 3G4, Canada, E-mail: xinmu@mpi-halle.mpg.de. https://orcid.org/0000-0001-6188-7969; and \textbf{Wesley D. Sacher}, Max Planck Institute of Microstructure Physics, Weinberg 2, 06120 Halle, Germany, E-mail:
wesley.sacher@mpi-halle.mpg.de.}
\affil[2]{Max Planck Institute of Microstructure Physics, Weinberg 2, 06120 Halle, Germany}
\affil[3]{University of Toronto, 10 King’s College Road, Toronto, Ontario, M5S 3G4, Canada}
\affil[4]{Advanced Micro Foundry Pte. Ltd., 11 Science Park Road, Singapore Science Park II, 117685, Singapore}
\affil[5]{ams OSRAM International GmbH, Leibnizstrasse 4, 93055 Regensburg, Germany}

\title{Hybrid integration of InGaN lasers in a foundry-fabricated visible-light photonics platform}

\runningauthor{X. Mu et al.}
\runningtitle{Hybrid-integrated InGaN lasers in a visible-light Si photonics platform}

\abstract{

Visible-spectrum photonic integrated circuits (PICs) present compact and scalable solutions for emerging technologies including quantum computing, biosensing, and virtual/augmented reality. Realizing their full potential requires the development of scalable visible-light-source integration methods compatible with high-volume manufacturing and capable of delivering high optical coupling efficiencies. Here, we demonstrate passive-alignment flip-chip bonding of 450-nm InGaN laser diodes onto a foundry-fabricated visible-light silicon (Si) photonics platform with silicon nitride (SiN) waveguides, thermo-optic (TO) devices, and photodetectors. Hybrid laser integration is realized using a sub-micron-precision die bonder equipped with a vision alignment system and a heatable pickup tool, allowing independent placement of multiple lasers onto a single Si chip. Co-design of the lasers and Si photonics, with lithographically defined alignment marks and mechanical stoppers, enables precise post-bonding alignment. Efficient optical coupling between lasers and the SiN waveguides is demonstrated, with a minimum measured coupling loss of 1.1 dB. We achieve a maximum on-chip optical power of 60.7 mW and an on-chip wall-plug efficiency of 7.8\%, the highest reported for hybrid-integrated visible-spectrum lasers, to our knowledge. An active PIC is also shown, integrating a bonded laser, an on-chip photodetector for power monitoring, and a thermo-optic switch for optical routing and variable attenuation. Overall, this work highlights passive-alignment flip-chip bonding as a practical, high-performance approach for integrating lasers onto visible-spectrum PICs. We envision that continued refinement of this technique within our photonics platform will support increasingly complex PICs with integrated lasers spanning the visible spectrum.
}

\keywords{laser integration; flip-chip bonding; visible-light integrated photonics; silicon photonics}


\maketitle

\section{Introduction} 

Silicon (Si) photonics in the visible spectrum represents a new frontier for photonic integrated circuits (PICs), broadening the range of applications beyond that of conventional Si photonics in the telecommunications O- and C-bands \cite{porcel2019silicon,sorace2019versatile,sacher2023active,Buzaverov2024}. Extending Si photonics into the visible spectrum unlocks opportunities in emerging technologies, including PIC-based microdisplays for virtual/augmented reality \cite{liu2023chip,shi2025flat,geuzebroek2025small}, quantum computing \cite{isichenko2023photonic,hattori2024integrated}, biosensing \cite{kumar2021analysis}, neurotechnology \cite{Mohanty2020,roszko2025foundry,Lakunina2025}, as well as free space and underwater optical communications \cite{sabouri2018design,notaros2023liquid}. To date, substantial progress has been made in establishing the fundamental on-chip components, such as passive silicon nitride (SiN) and aluminum oxide waveguide devices \cite{sacher2019visible,sorace2019versatile,blasco2024silicon}, beam scanners \cite{ChulShin2020,ankita2025}, optical switches and modulators \cite{liang2021robust,dong2022piezo,notaros2022integrated,govdeli2025integrated}, and photodetectors \cite{Yanikgonul2021,de2022amorphous,alperen2025}. Despite these advances, realizing the full potential of visible-light Si photonics will require integrated light sources that collectively span the visible spectrum. 

A key challenge in integrating lasers onto Si photonic platforms is the poor light-generation efficiency of the standard waveguide materials: Si has an indirect bandgap, and the SiN used here is an amorphous, wide-bandgap dielectric. Integration with direct-bandgap semiconductors is therefore essential for efficient on-chip light sources. Various laser integration strategies have been investigated, including wafer bonding \cite{liang2021recent}, direct epitaxial growth \cite{kum2019epitaxial}, micro-transfer printing \cite{roelkens2022micro}, and flip-chip bonding \cite{theurer2020flip}. With most prior work focused on telecommunications wavelengths using indium phosphide gain media, visible-spectrum integration requires multiple gain materials, such as indium gallium nitride (InGaN) for blue/green and gallium arsenide-based compounds for red, along with wavelength-specific optimization of device compositions. Hybrid integration of pre-fabricated laser diodes (LDs) offers a practical route to incorporating visible light sources into Si photonics, accommodating the diverse materials and wide wavelength range of the visible spectrum. Recent work has explored actively aligning edge-emitting LDs to Si photonic chip facets via butt-coupling \cite{franken2021hybrid,corato2023widely}, achieving precise alignment and high coupling efficiency but relying on slow, feedback-based assembly and limiting integration density due to facet and LD submount size constraints. These factors bottleneck assembly throughput for volume production and restrict PIC complexity. 

An alternative to active-alignment techniques for hybrid laser integration is passive-alignment flip-chip bonding, a method that has been widely demonstrated in conventional infrared Si photonics \cite{matsumoto2018hybrid,theurer2019flip,marinins2022wafer}. Here, etched recesses (LD trenches or bonding sockets) are defined on the Si photonic platform. Lithographically-defined alignment marks on both the LDs and Si chips enable precise in-plane LD positioning, while mechanical stoppers on both dies, often formed using etch-stop layers, ensure accurate vertical self-alignment via contact between corresponding features. Electrical connection is achieved by interfacing the LD metal contacts with solder bumps and under-bump metallization (UBM) pads on the Si chips. Sub-micron post-bonding alignment of LDs is typically achieved using commercial die bonders. These systems are widely used in microelectronics packaging, and existing models are capable of high-throughput assembly. By combining precisely defined on-chip alignment features with advanced die-bonding tools, passive-alignment flip-chip bonding offers a scalable route for hybrid laser integration, enabling high coupling efficiency, high throughput assembly, and compatibility with multiple laser types. It also supports pre-screening and burn-in testing of LDs to identify known-good dies, reducing the impact of LD yield on packaged PIC yield. While proven in infrared Si photonics, applying this approach to visible-light platforms is more challenging due to the smaller optical mode size of LDs at shorter wavelengths, which tightens alignment tolerances. Consistent with this, recent demonstrations of visible-spectrum LDs flip-chip bonded onto PICs have thus far shown limited coupled optical powers. Reference \citenum{kluge2022flip} reported a PIC-based external cavity laser with a flip-chip bonded red gain chip, although lasing was not achieved in the prototype. Reference \citenum{geuzebroek2025small} demonstrated red, green, and blue flip-chip bonded lasers on a SiN PIC, with a combined maximum output power of 10 mW.

In this work, we demonstrate the hybrid integration of 450-nm InGaN laser diodes into a foundry-fabricated visible-light Si photonics platform via passive-alignment flip-chip bonding, using localized heating compatible with bonding multiple LDs per Si chip. The platform integrates SiN nanophotonic waveguides, thermo-optic tuners, and waveguide-coupled Si photodetectors \cite{sacher2023active,Bebeti2025CLEO}. From 40 bonded samples, we achieve a maximum on-chip optical power of 60.7 mW and wall-plug efficiency (referenced to on-chip power) of up to 7.8\%. To the best of our knowledge, this power is a record among visible-spectrum hybrid-integrated lasers. The devices maintain stable operation at temperatures up to 80 $\degree$C, and statistical analysis confirms sub-micron post-bonding alignment within the placement accuracy of our die bonder. Although this accuracy constrains assembly yield, we outline pathways for improvement. We also present a PIC integrating a flip-chip bonded LD, a monolithically integrated photodetector, and a thermo-optic switch, demonstrating the platform's core active functions. Overall, this work establishes a scalable platform for laser integration in visible-light Si photonics, paving the way for advanced active PICs with arrays of hybrid-integrated visible lasers. Leveraging foundry fabrication and high-throughput flip-chip assembly, the approach is poised to scale in complexity and manufacturing volume, enabling PIC-based solutions for emerging visible-spectrum applications. An initial report of these results was presented in our conference abstract, Ref. \citenum{mu2025flip}.

\begin{figure*}[h]
\centering
\includegraphics[width=0.8\textwidth]{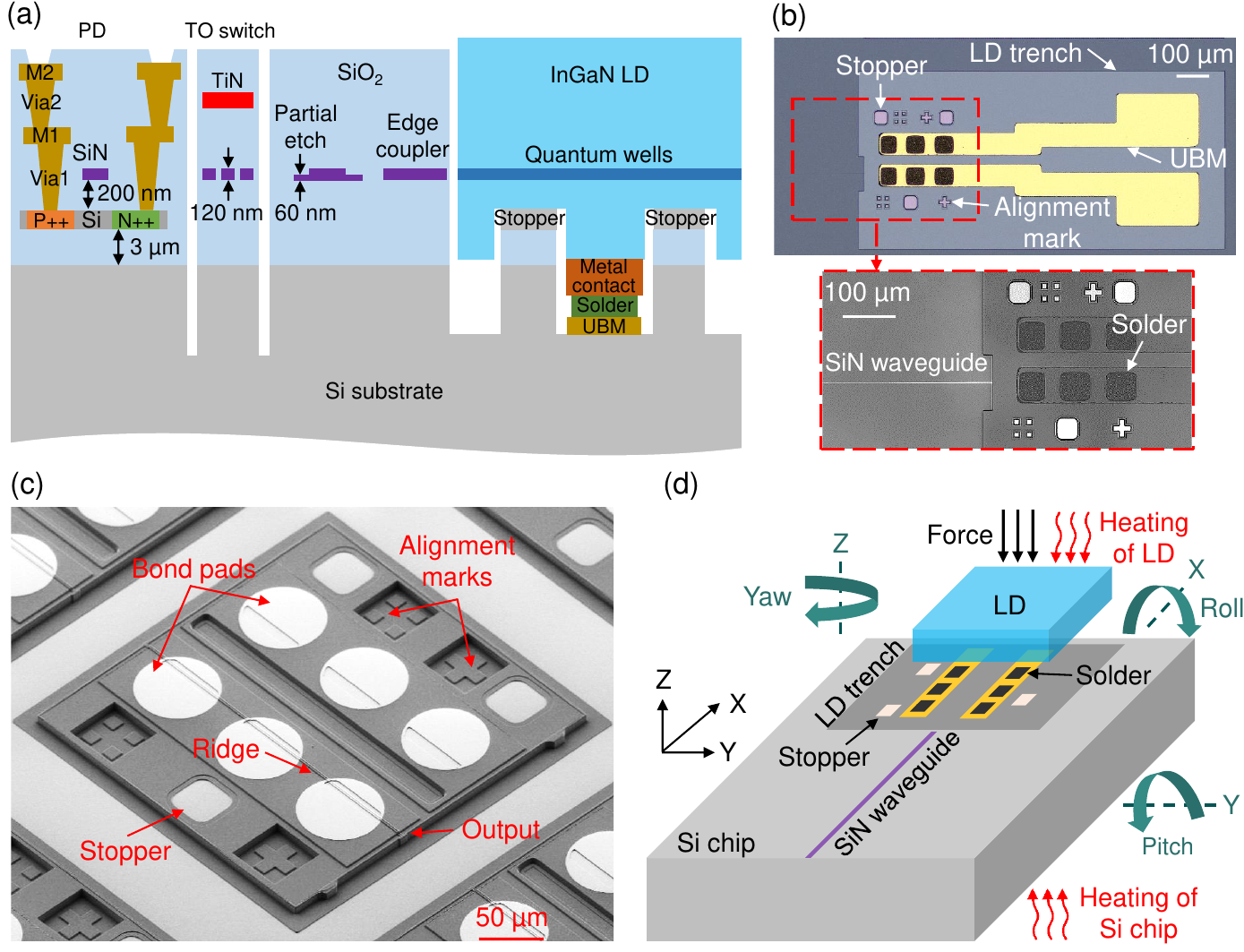}
\caption{\justifying 
Hybrid integration of InGaN laser diodes (LDs) in a visible-light Si photonics platform. (a) Cross-section of the photonics platform. PD: photodetector, M: metal, SiN: silicon nitride, TO switch: thermo-optic switch, TiN: titanium nitride, UBM: under-bump metallization. (b) (Top) Optical micrograph of an LD bonding socket on a Si photonic chip, and (bottom) confocal laser scanning micrograph of the socket with improved visibility of the SiN waveguide. (c) Scanning electron micrograph of the top surface of an InGaN LD before singulation. (d) Conceptual illustration of the flip-chip LD bonding method, with an LD close to the socket. The definition of the coordinate system and rotation axes is given.
\label{fig1}}
\end{figure*}

\section{Design and fabrication} 

Figure \ref{fig1}(a) illustrates the schematic of the visible-light Si photonics platform, featuring SiN waveguides, thermo-optic switches, waveguide-coupled photodetectors, and structures for hybrid integration of laser diodes. The platform was fabricated at Advanced Micro Foundry on 200-mm diameter silicon-on-insulator (SOI) wafers \cite{govdeli2024broadband}. The fabrication process began with the formation of 220-nm thick Si photodetectors, including Si patterning and doping steps. Subsequently, SiN waveguides were defined via plasma enhanced chemical vapor deposition, 193-nm deep ultraviolet photolithography, and reactive ion etching. Full- and partial-etch steps were used for definition of 120-nm and 60-nm-thick waveguides. Each Si PIN-junction photodetector was located beneath a SiN waveguide, allowing evanescent coupling of guided light from the SiN waveguide into the underlying Si photodetector. Next, metallization steps were performed with two aluminum routing layers and interlayer vias, as well as the titanium nitride (TiN) resistive heaters. Chemical mechanical polishing was used for layer planarization. SiO$_2$ and Si etch steps defined the flip-chip bonding sockets (LD trenches), mechanical stoppers (for vertical positioning), and alignment marks (for in-plane alignment), Figure \ref{fig1}(b). The Si device layer of the SOI wafers served as both the top surface of the mechanical stoppers and as an etch stop during stopper patterning, thereby minimizing height variability and enabling precise vertical alignment between the stoppers and the waveguides. Under-bump metallization and gold-tin (AuSn) solder bumps were patterned within the bonding sockets to interface with the metal contacts (bond pads) of LDs for electrical connectivity. Two types of solder bump patterns were implemented --- rows of discrete bumps in Figure \ref{fig1}(b) and continuous stripes in Figure \ref{fig7}(a). The LD trench configuration enables direct thermal contact between the LD and Si substrate through thin solder and UBM layers, enabling efficient heat sinking. The fabrication process concluded with etching of deep Si trenches to form the chip facets, followed by wafer dicing. 

In parallel, blue ($\approx$450 nm) edge-emitting, single-spatial-mode, InGaN laser diodes were fabricated by ams OSRAM \cite{nahle2025options}. Each laser diode chip was $\approx$100 $\upmu$m thick with a footprint of $\approx$400 $\upmu$m $\times$ 400 $\upmu$m, Figure \ref{fig1}(c). The InGaN LDs were co-designed in conjunction with the bonding sockets on the Si photonic platform, specifically for hybrid laser integration with passive-alignment flip-chip bonding. Mechanical stopper trenches and alignment marks were formed on the LDs by trench etching, matching counterparts on the Si photonic chips. The LDs featured top-side P and N metal contacts, allowing for electrical interfacing with the pre-patterned solder bumps and UBM pads on the Si chips. Precise lithography and etching of the LDs and Si chips ensured that the alignment marks and stoppers provided accurate in-plane and vertical positioning during placement. SiN edge couplers, located at the facets of the bonding sockets on the Si photonic chips, coupled the emitted light from the flip-chip bonded lasers into the SiN waveguides, Figure \ref{fig1}(d). The detailed laser-to-waveguide coupler design is discussed in Section \ref{coupling}.

\section{Hybrid integration method} 
\label{bonding method}

\subsection{Flip-chip bonding process}
\label{process}

\begin{figure}[t!]
\includegraphics[width=\columnwidth]{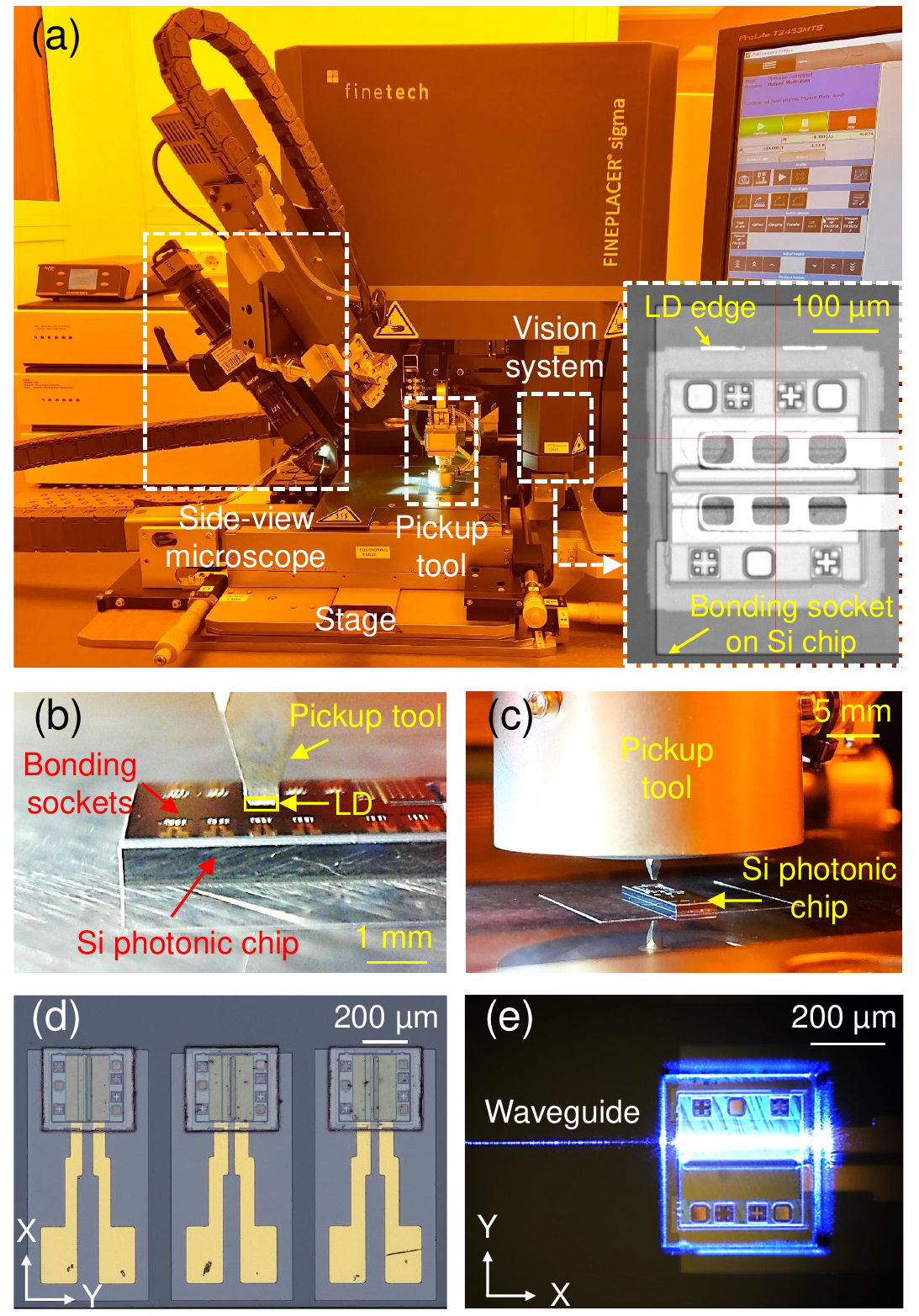}
\caption{\justifying
Flip-chip bonding technique. (a) Photograph of the die bonder with a vision alignment system, side-view microscope, and heatable pickup tool. Inset: overlaid view from the vision system of an LD and a bonding socket prior to bonding. (b) Micrograph of an LD held by the pickup tool approaching a Si photonic chip; the image was captured with a second side-view microscope not shown in (a). (c) Photograph of the pickup tool above a Si photonic chip. (d) Micrograph of three flip-chip bonded LDs on one Si photonic chip. (e) Micrograph of a bonded LD with emitted light coupled to a SiN waveguide.
\label{fig2}}
\end{figure}

Hybrid laser integration via passive-alignment flip-chip bonding was implemented using a semi-automated die bonder with sub-micron placement accuracy (Fineplacer sigma, Finetech), Figure \ref{fig2}(a). The system picked up the laser diode using a vacuum pickup tool and performed an aligned touchdown onto a bonding socket on the Si photonic chip, applying a controlled sequence of pressure and heat for thermocompression bonding, as conceptually illustrated in Figure \ref{fig1}(d). The die bonder featured a vision alignment module that overlaid live microscope images of the laser diode (held by the vacuum pickup tool) and the Si photonic chip (mounted on the translation stage of the die bonder), enabling visualization of the relative positions of cross-shaped alignment marks on both chips [inset, Figure \ref{fig2}(a)]. A side-view microscope facilitated real-time monitoring of the LD touchdown onto the Si photonic chip. Figure \ref{fig2}(b) shows the side view of the LD approaching the Si photonic chip, taken with a microscope different from the side-view microscope of the die bonder. The heatable vacuum pickup tool, Figure \ref{fig2}(c), allowed localized heating of the LD to selectively melt solder bumps only within the targeted bonding socket. This selective heating enabled sequential integration of multiple LDs onto a single Si chip, Figure \ref{fig2}(d). Supplemental heat was applied to the Si photonic chip through the die bonder alignment stage, reducing the temperature required by the pickup tool to reach the solder melting point, which in turn minimized thermal exposure to the LD.  

The detailed flip-chip bonding process began with securing the Si photonic chip onto the translation stage of the die bonder using vacuum suction. Next, an InGaN laser diode, stored with contact-surface down in a nano device tray (NDT-XT-RA, Gel-Pak), was picked up by the heatable vacuum pickup tool. Both the Si photonic chip and the LD were then pre-heated to 270 \degree C, a temperature below the melting point of the AuSn solder. This step minimized the temperature difference between the two chips in the subsequent bonding process, during which the pickup tool was heated up to 400 $\degree$C, thereby reducing alignment offsets caused by thermal expansion. The target bonding socket on the Si photonic chip was positioned beneath the LD using the translation stage. Precise alignment was achieved by manually centering the alignment marks on both the LD and the Si photonic chip using the vision system. Once aligned, the LD was lowered onto the bonding site, and the thermal bonding profile was initiated upon detecting a touchdown force of 2 N. During the bonding process, the Si chip temperature was maintained at 270 \degree C, while the pickup tool was heated to a peak temperature of 400 \degree C at a ramp rate of 5 \degree C/s. After heating at 400 \degree C for 30 s, both the translation stage and the pickup tool were cooled to 40 \degree C at a rate of 5 \degree C/s using ambient air delivered through vent holes in the stage. A constant touchdown force of 2 N was maintained throughout the process to secure the LD in position and ensure consistent mechanical contact with the bonding socket on the Si photonic chip. Following cooling and solidification of the solder bumps, vacuum suction from the pickup tool was released, and the tool was retracted.

Because of the tapered profile and small tip of the pickup tool, and contact with the lower-temperature Si chip, the LD temperature during bonding was expected to remain below the 400 \degree C tool setting. The peak temperature was estimated at $\approx$315 \degree C, between the pickup tool (400 \degree C) and stage (270 \degree C), measured using a thermometer (Voltcraft, 302K/J) with a thermocouple (of similar dimensions to the LD) placed between them. These measurements, combined with iterative adjustment of bonding parameters including force, ramp rate, and duration, informed process development aimed at achieving consistent post-bonding placement accuracy and solder joint integrity. A complete flip-chip bonding cycle, including sample loading, LD pickup, alignment, thermocompression bonding, and cooling, required $\approx$15 min. Figure \ref{fig2}(e) shows a hybrid-integrated LD following the flip-chip bonding procedure, with blue emitted light coupled into a SiN photonic waveguide via an edge coupler. Detailed characterization of the LDs is presented in Section \ref{LD test}.

\subsection{Simulated alignment sensitivity}
\label{coupling}

\begin{figure}[t!]
\centering
\includegraphics[width=0.8\columnwidth]{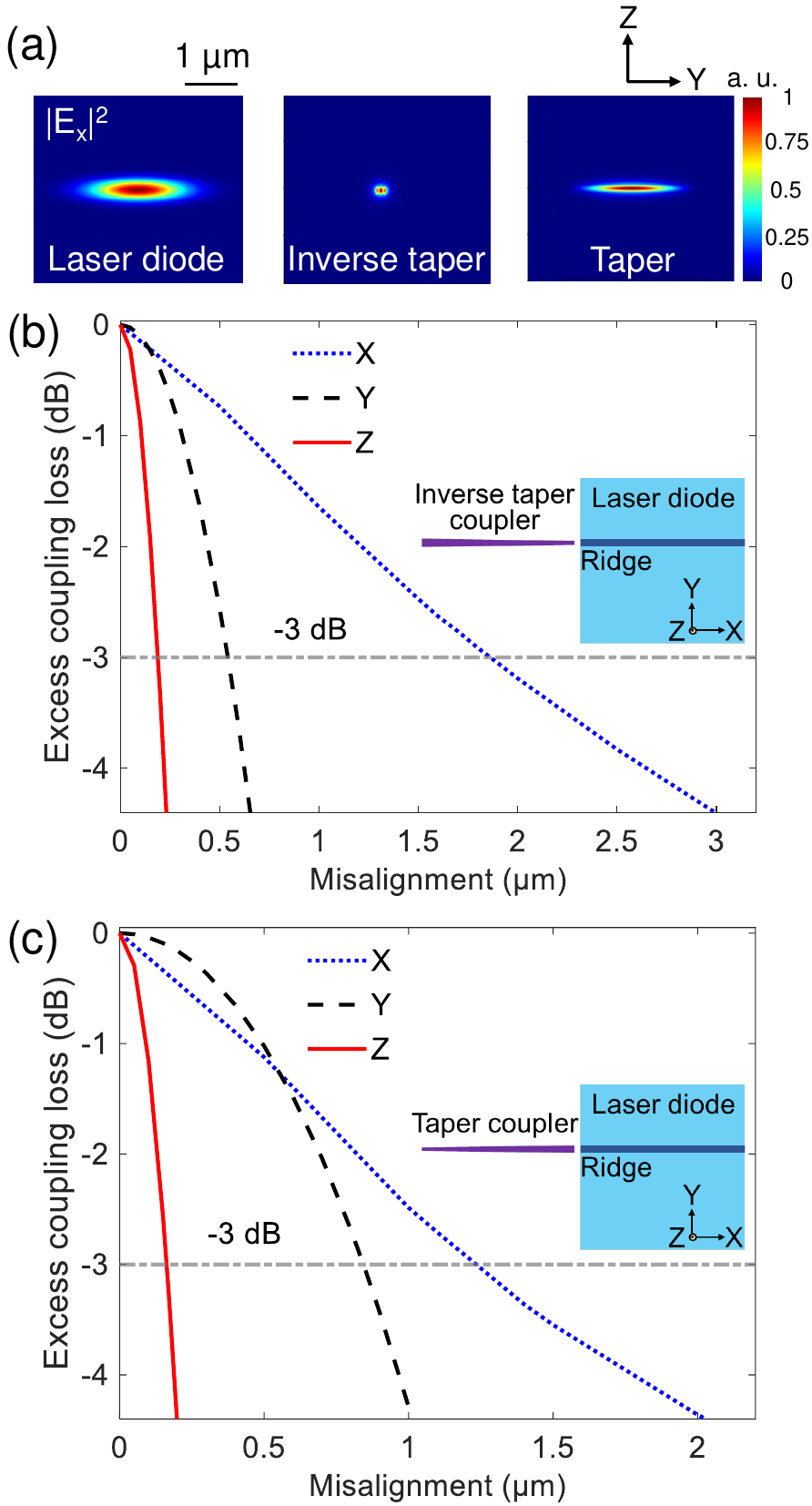}
\caption{\justifying
Simulated laser-to-waveguide coupling. (a) Optical mode profiles at the laser facet and SiN coupler facet (both inverse taper and taper design); $\lambda = $ 450 nm, transverse-electric (TE) polarization. Simulated in-plane misalignment tolerance of (b) inverse taper and (c) taper couplers. Insets: Illustrations of the LD-to-waveguide coupling (top-down view).
\label{fig3}}
\end{figure}

To enable efficient optical coupling between the flip-chip bonded laser diodes and Si photonic chips, two types of LD-to-waveguide couplers were evaluated. The first design, a SiN inverse taper coupler, begins with a 150-nm-width tip at the bonding socket facet, expanding to a 400-nm-wide single-mode waveguide over a 150-$\upmu$m length. The second design, a SiN taper coupler, narrows from a 3-$\upmu$m width at the socket facet to 400 nm over a length of 190 $\upmu$m. The inverse taper is implemented in full-thickness (120 nm) SiN, while the taper coupler uses the partially-etched 60-nm-thick SiN layer. Figure \ref{fig3}(a) presents simulated optical mode profiles at the LD emission facet and the coupler facets, obtained using a finite-difference eigenmode solver (Ansys Lumerical). The coupler dimensions were chosen to optimize mode matching and coupling efficiency with the LD output. These dimensions were selected within the feature size constraints of the fabrication process. As part of the design, the taper coupler was implemented in the partially-etched SiN to reduce vertical (Z-axis) optical confinement.

Modal overlap calculations were performed to evaluate laser-to-waveguide coupling efficiency under ideal alignment conditions as well as with transverse [Y- and Z-axis, Figures \ref{fig1}(d) and \ref{fig3}(a)] misalignment. Additional three-dimensional finite-difference time-domain (FDTD) electromagnetic simulations (Ansys Lumerical) were performed to assess the impact of longitudinal (X-axis) misalignment. With perfect alignment, the simulated coupling losses for the inverse taper and taper couplers are 2.9 dB and 0.6 dB, respectively. Excess coupling losses due to misalignment, relative to the perfectly aligned case, are shown in Figures \ref{fig3}(b) and \ref{fig3}(c).

Table \ref{tab:Table1} summarizes the 3-dB misalignment tolerances between the flip-chip bonded LD and the SiN couplers on the Si photonic chip. Since the laser mode is more confined in the vertical direction (Z-axis), optical coupling is more sensitive to vertical misalignment compared to the other two directions. Relative to the inverse taper, the taper coupler shows increased sensitivity to misalignment along the X- and Z-axes, which is attributed to its reduced vertical mode size. By contrast, the taper coupler is more tolerant to Y-axis misalignment because of its larger horizontal mode width. Overall, the taper coupler achieves lower coupling loss from simulation due to improved mode matching with the LD, while the inverse taper design offers greater tolerance to misalignment. For both coupler geometries, the misalignment tolerances are smaller than those typically reported for near-infrared hybrid-integrated lasers \cite{theurer2020flip,marinins2022wafer}, primarily due to the shorter wavelength ($\approx$450 nm) and the correspondingly smaller optical mode size of the blue InGaN LD. The coupler design may be further optimized to improve misalignment tolerance, for example, through the use of multi-tip geometries \cite{kluge2022flip} or adiabatic couplers \cite{li2025novel}. Additional simulation results for the rotational misalignment tolerances of both coupler designs are provided in Figure \ref{figS1} in the Supplementary Material.

\begin{table} [h]
\caption{Summary of simulated 3-dB misalignment tolerances of the LD-to-waveguide couplers along the X, Y, and Z axes. Unit: $\upmu$m.}
\centering
\begin{tabular}{w{c}{2.5cm} w{c}{1cm} w{c}{1cm} w{c}{1cm}}
Coupler design 	& X 	& Y & Z	\\ \midrule
Inverse taper & 1.87 & 0.55 & 0.20			\\
Taper & 1.23 & 0.85 & 0.16			\\
\end{tabular}
\label{tab:Table1}
\end{table}

\subsection{Bonding process characterization}
\label{bonding performance}

\begin{figure*}[h]
\centering
\includegraphics[width=0.67\textwidth]{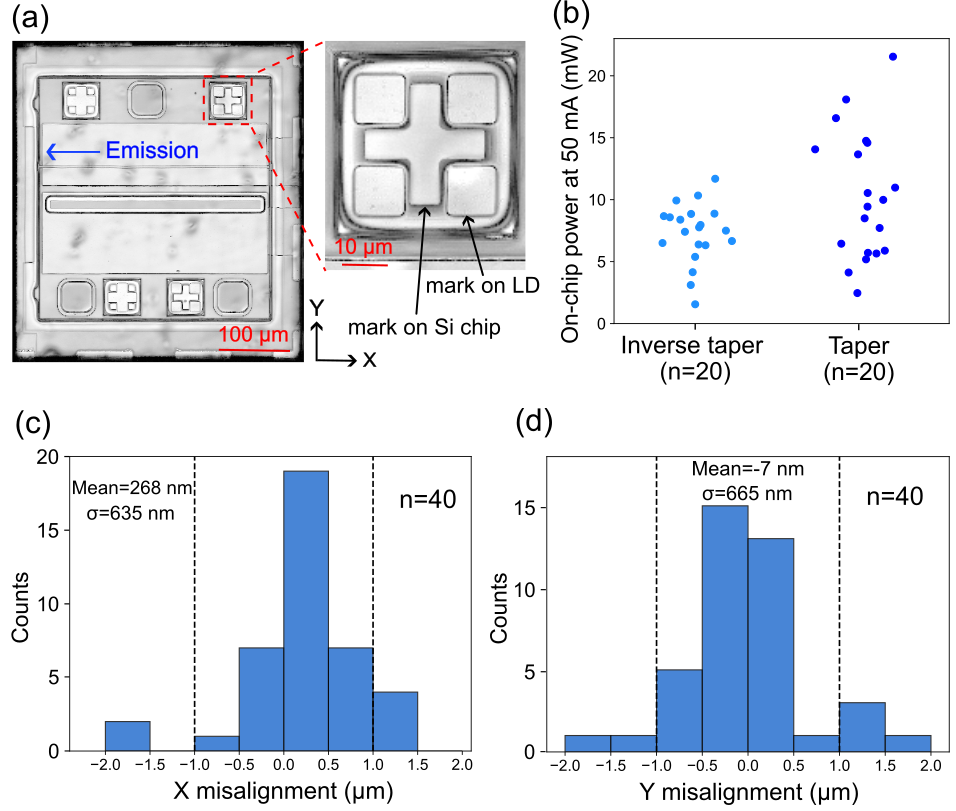}
\caption{\justifying
Characterization of the flip-chip bonding process. (a) Confocal laser scanning micrograph of a representative flip-chip bonded LD (left) with an enlarged view of one set of alignment marks (right). Cross: Si photonic chip mark; squares: LD mark; both visible through the LD. Micrograph quality was limited by imaging through the $\approx$100-$\upmu$m-thick LD substrate. (b) On-chip optical power distribution of flip-chip bonded LDs (at 50-mA drive current) with inverse taper and taper laser-to-waveguide couplers; n=20 samples for each coupler design. Histograms of in-plane misalignment after flip-chip bonding along the (c) X- and (d) Y-axis; n=40, bin size=0.5 $\upmu$m, both coupler designs included.
\label{fig4}}
\end{figure*}

To assess the repeatability of the flip-chip bonding process, 40 laser bonding trials were performed (20 trials for each of the two LD-to-waveguide coupler designs). For each bonded LD, the post-bonding in-plane misalignment and the on-chip optical power coupled into the SiN waveguide were measured. Due to the transparency of the InGaN LD substrate, the alignment marks on the LD and Si photonic chip remained visible following bonding, Figure \ref{fig4}(a). This property enabled in-plane misalignment measurements via confocal laser scanning microscopy using a 3D surface profiler (Keyence VK-X3000) to image the alignment marks. For each sample, the in-plane misalignment was calculated as the average offset across the four alignment mark pairs.

Histograms of the post-bonding in-plane misalignment along the X- and Y-axes for all 40 samples are presented in Figures \ref{fig4}(c) and \ref{fig4}(d), respectively. The measured misalignment exhibited standard deviations ($\sigma$) of 635 nm (X-axis) and 665 nm (Y-axis), with corresponding mean values of 268 nm and -7 nm, respectively. Misalignment within $\pm$1 $\upmu$m along both axes was observed for 85\% of the bonded samples. The confocal laser scanning microscopy system exhibited a pixel-resolution-limited measurement precision of $\approx$$\pm$140 nm, constrained by the reduced optical resolution from imaging through the $\approx$100-$\upmu$m-thick InGaN LD substrate. The post-bonding laser-to-chip alignment precision was mainly limited by the placement accuracy of the die bonder. Furthermore, the use of a die bonder with a chip stage translation resolution of 1 $\upmu$m introduced additional alignment constraints.

\begin{table} [h]
\caption{Summary of measured rotational misalignment of the flip-chip bonded samples. The maximum angular tilt is reported. The definition of rotation axes is given in Figure \ref{fig1}(d).}
\centering
\begin{tabular}{w{c}{2.5cm} w{c}{1cm} w{c}{1cm} w{c}{1cm}}
Coupler design 	& Pitch 	& Roll & Yaw	\\ \midrule
Inverse taper & 0.05$\degree$ & 0.05$\degree$ & 0.12$\degree$			\\
Taper & 0.07$\degree$ & 0.03$\degree$ & 0.20$\degree$			\\
\end{tabular}
\label{tab:Table2}
\end{table}

The angular tilt of the flip-chip bonded LDs was characterized using white light interferometry with the 3D surface profiler (see Figure \ref{figS2} in the Supplementary Materials). The vacant area adjacent to the bonding socket on the Si photonic chip served as the reference plane for the tilt measurements. The measured angular tilts of the flip-chip bonded LD samples with both coupler designs are summarized in Table \ref{tab:Table2}. According to the simulation results in Figure \ref{figS1}, these angular deviations correspond to an estimated excess coupling loss of $<$0.5 dB. Overall, post-bonding rotational misalignment was minimal, as the alignment marks and mechanical stoppers effectively constrained angular tilt during flip-chip bonding.

For characterization of the optical power, laser light coupled into each SiN waveguide was routed to an edge coupler at an output chip facet, coupled to a single-mode optical fiber (Nufern S405-XP), and measured with an optical power meter (Newport 2936-R, 818-SL/DB). The on-chip optical power was calculated by subtracting the fiber-to-chip edge coupler loss (measured separately using dedicated test structures) from the measured off-chip optical power. LDs bonded onto Si chips were characterized with the chips mounted on a copper heat sink atop a thermoelectric cooler (TEC) maintained at 21 \degree C unless otherwise specified. Each LD was driven by a source measure unit (SMU, Keysight, B2912A), with current supplied through tungsten DC probes contacting the UBM pads on the Si photonic chip.

Figure \ref{fig4}(b) summarizes the measured on-chip optical power from the 40 flip-chip bonded LD samples. 85\% of the LD samples with inverse taper couplers and 90\% of those with taper couplers exhibited on-chip optical power $\ge$5 mW. LD-to-waveguide coupling loss was determined by comparing the on-chip optical power of flip-chip-bonded LDs (pulsed operation) with emission power of a reference LD (no flip-chip process, pulsed operation) at identical drive currents. Pulsed operation reduced the effects of differing heat sinking and also reduced the resulting variations in operating temperature between the flip-chip-bonded and reference LDs, enabling accurate comparison. Twelve samples were selected for these additional measurements. The minimum laser-to-waveguide coupling loss was $\approx$4.1 dB for inverse taper couplers and $\approx$1.1 dB for taper couplers, corresponding to excess coupling losses (due to misalignment) of $\approx$1.2 dB and $\approx$0.5 dB, respectively (see Table \ref{tab:Table3}). The on-chip power distribution in Figure \ref{fig4}(b) is consistent with the simulation results in Section \ref{coupling}, confirming that, relative to inverse taper couplers, taper couplers exhibit higher sensitivity to misalignment (indicated by a wider spread in measured power) but achieve higher coupling efficiency (reflected by higher maximum optical power levels).

Table \ref{tab:Table3} summarizes the simulated and measured coupling losses together with the on-chip optical power. The measured minimum coupling losses agree with the simulated results within 1.2 dB for the inverse taper and 0.5 dB for the taper, with the larger discrepancy for the inverse taper possibly arising from its increased sensitivity to small variations in tip width. The on-chip optical power across samples showed significant variability, attributed to the LD-to-waveguide misalignment tolerance, which is comparable to the placement accuracy of the die bonder. The measured standard deviation of post-bonding alignment along the X-axis (635 nm) is 2.9$\times$ smaller for the inverse taper and 1.9$\times$ smaller for the taper compared to the simulated 3-dB misalignment tolerance in Table \ref{tab:Table1}. Along the Y-axis, the standard deviation (665 nm) is 1.2$\times$ larger for the inverse taper and 1.3$\times$ smaller for the taper than the simulated tolerance. Although these values remain within the die bonder’s specified placement accuracy, reducing variability in on-chip optical power will require systems with higher precision. Section \ref{performance comparison} provides estimates of coupling efficiency improvements achievable with sub-300 nm precision, following Ref. \citenum{marinins2022wafer}.

\begin{table} [!ht]
\caption{Summary of simulated and minimum measured coupling losses, and measured on-chip optical power for both laser-to-chip coupler designs.}
\centering
\begin{tabular}{m{2cm} m{1.4cm} m{1.5cm} m{2cm}}
Coupler design 	& Simulated loss 	& Measured loss (min.)  & Measured on-chip power$^a$	\\ \midrule
Inverse taper & 2.9 dB & 4.1 dB	& 7.5 $\pm$ 2.4 mW		\\
Taper & 0.6 dB & 1.1 dB 	& 	10.5 $\pm$ 5.0	mW \\
\end{tabular}
\justifying
\footnotesize{$^a$Drive current of 50 mA, power represented by mean $\pm$ $\sigma$.}
\label{tab:Table3}
\end{table}

The flip-chip bonding process uses localized heating from the pickup tool to independently bond multiple LDs onto a single Si photonic chip, Figure \ref{fig2}(d). When sockets are closely spaced, successive bonding steps may thermally affect previously bonded LDs, potentially altering their alignment. To evaluate this possibility, we analyzed on-chip optical power as a function of bonding order. Of the 40 integrated LDs, 26 were bonded onto Si chips hosting multiple LDs from the sample set; their measurements are presented in Section \ref{multi-die} of the Supplementary Materials. 

For inverse taper couplers, the mean power of initially bonded LDs (exposed to thermal cycles from subsequent bonding steps) was 86\% of that of last-bonded LDs, while for taper couplers the corresponding value was 64\%. Comparison with Si chip samples on which only a single LD was bonded indicates that these mean values fall within the variability of the flip-chip process. Thus, while the data suggest a possible influence of bonding order, any effect is difficult to distinguish from process variability. Improvements to the flip-chip process and LD-to-waveguide coupler design for reduced coupling variability, may help clarify the significance of this effect, and increasing the bonding socket pitch could reduce thermal interactions between successive bonding steps.

\section{Hybrid-integrated laser diodes} 
\label{LD characterization}

\subsection{L-I-V curves and optical spectra} 
\label{LD test}

\begin{figure}[t]
\centering
\includegraphics[width=0.75\columnwidth]{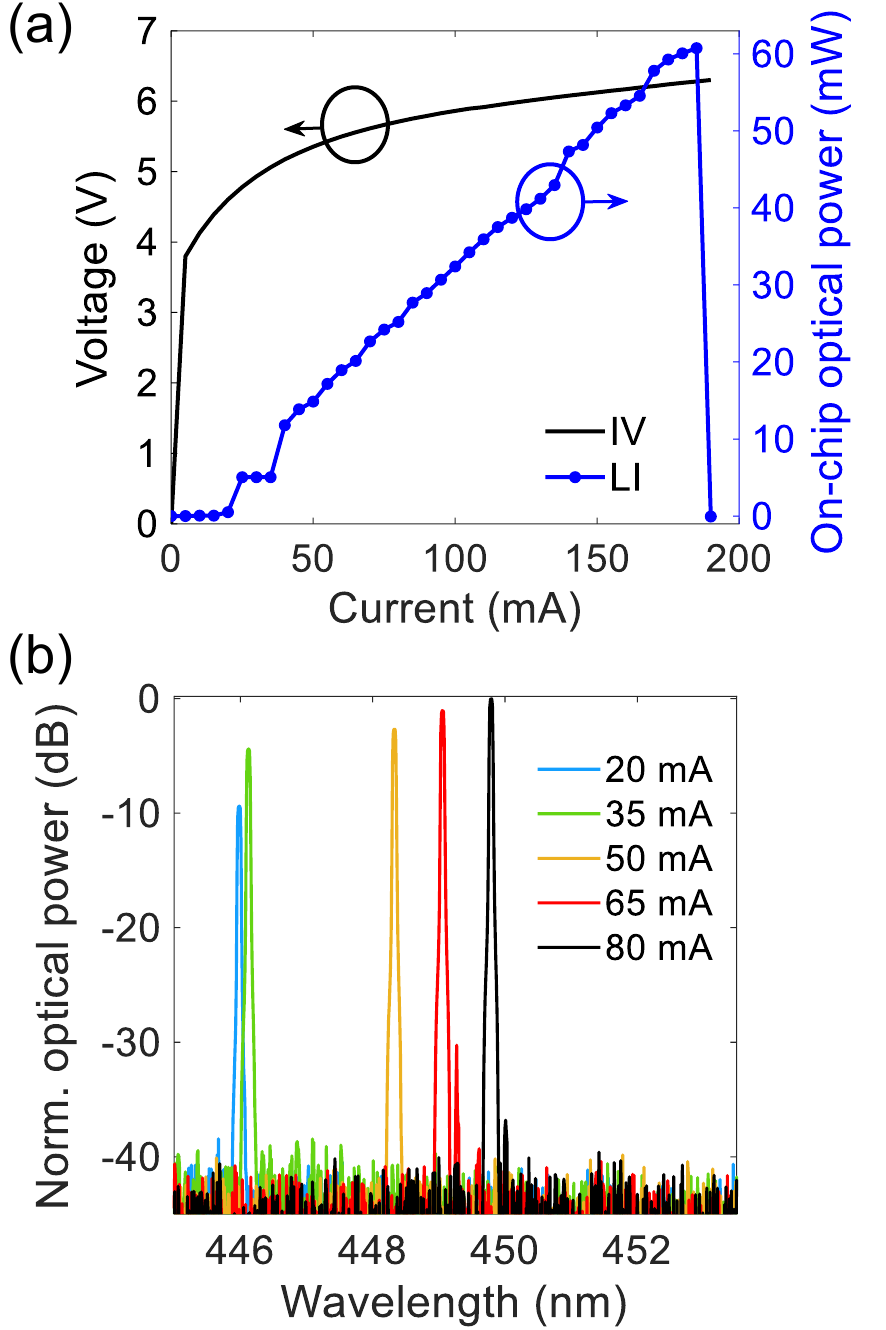}
\caption{\justifying
Characterization of a flip-chip-bonded laser diode showing the highest on-chip optical power achieved in this work. (a) L-I-V curves measured in continuous-wave (CW) mode. L: On-chip optical power; I: drive current; V: voltage drop. Device failure occurred at a drive current of 185 mA. (b) Optical spectra at different laser drive currents (normalized to the peak power at 80 mA current).
\label{fig5}}
\end{figure}

The hybrid-integrated laser diodes were characterized by measuring the optical power-current-voltage (L-I-V) curves and optical spectra under continuous-wave (CW) operation. Figure \ref{fig5}(a) shows the L–I–V characteristics of a bonded LD aligned to a taper coupler on a Si photonic chip. This device achieved the highest measured peak on-chip (SiN waveguide-coupled) optical power in this work, 60.7 mW, before device failure, which was attributed to the high current density at a drive current of 185 mA. The maximum on-chip wall-plug efficiency (WPE), defined with respect to the on-chip optical power, was $\approx$6.0\% at a drive current of 150 mA and decreased to $\approx$5.3\% at 185 mA for this sample. Figure \ref{figS4} presents additional L-I-V curves from two other LD samples (LD1 and LD2), where thermal rollover (without LD damage) was observed with drive currents exceeding 144 mA and 132 mA, respectively. LD1 and LD2 reached maximum on-chip optical powers of 42.1 mW and 46.8 mW, respectively, with LD2 exhibiting the highest on-chip WPE ($\approx$7.8\%) among the 40 bonded LD samples.

Improvements in the on-chip optical power may be realized through post-assembly epoxy underfill, which decreases the large refractive index discontinuity in the gap between the LD and SiN waveguide coupler, reducing reflections at the optical coupling interface. Additional improvement in optical power may be achieved through enhanced heat dissipation, for example, by employing thermally conductive epoxy encapsulation and/or larger laser bond pads and solder bumps. In this work, additional tests were performed to evaluate the effects of epoxy underfill and encapsulation. Polydimethylsiloxane (PDMS, Sylgard 184) was applied at the laser-to-waveguide coupling interface, followed by a layer of thermally conductive epoxy (Cotronics, Duralco 128) on top of the LD to promote heat dissipation (see Section \ref{epoxy} in the Supplementary Material). Across five samples, a moderate average on-chip optical power increase of 12.6\% was observed. Epoxy underfill and encapsulation were applied only in the tests shown in Figure \ref{figS5}; unless otherwise specified, LD characterization was performed without these steps.

Figure \ref{fig5}(b) presents optical spectra of the flip-chip-bonded LD with the highest on-chip optical power, measured at different drive currents with an optical spectrum analyzer (Yokogawa AQ6374). Emission peaks were centered near 450 nm, exhibiting a red shift with increasing drive current, consistent with laser heating at elevated drive currents. The measured spectral linewidth (full width at half maximum) remained within 0.05 - 0.10 nm, limited by the 0.05 nm resolution of the optical spectrum analyzer. At drive currents of 65 mA and 80 mA, multiple longitudinal modes were present, with side-mode suppression ratios exceeding 29 dB.

Additional tests were conducted to examine the polarization properties of the LDs and to assess the impact of the flip-chip bonding process on LD performance. Polarization was analyzed for both a flip-chip-bonded and a bare LD by sending the emitted light (from the Si photonic chip or bare LD) through a polarizer onto a free-space detector (Newport 918D-ST-SL). Both LD samples were measured to be TE-polarized, with polarization extinction ratios $>$20 dB. Separately, the potential impact of the flip-chip bonding process (heat and pressure) on the performance of the LD chips was evaluated by comparing the series resistance and emission power of the LDs before and after exposure to heat and pressure from the die bonder (see Section \ref{bonding impact} in the Supplementary Material). To enable direct measurement of LD emission power (rather than waveguide-coupled power), the LDs were mounted contact-side-up on a blank Si test chip. Heat and pressure were then applied to the LD top surfaces via the pickup tool, with additional heating from the die bonder stage, following the same force and temperature profiles as in the full flip-chip bonding process. Only minimal changes in LD performance were observed; averaged across three samples, the series resistance increased by $\approx$13.6\% and the emission power increased by $\approx$13.3\%. This slight increase in emission power may result from curing of the metal adhesive attaching the LD to the Si chip and the applied pressure, which together may have influenced thermal conduction.

\subsection{Temperature dependence} 

\begin{figure}[!b]
\centering
\includegraphics[width=0.8\columnwidth]{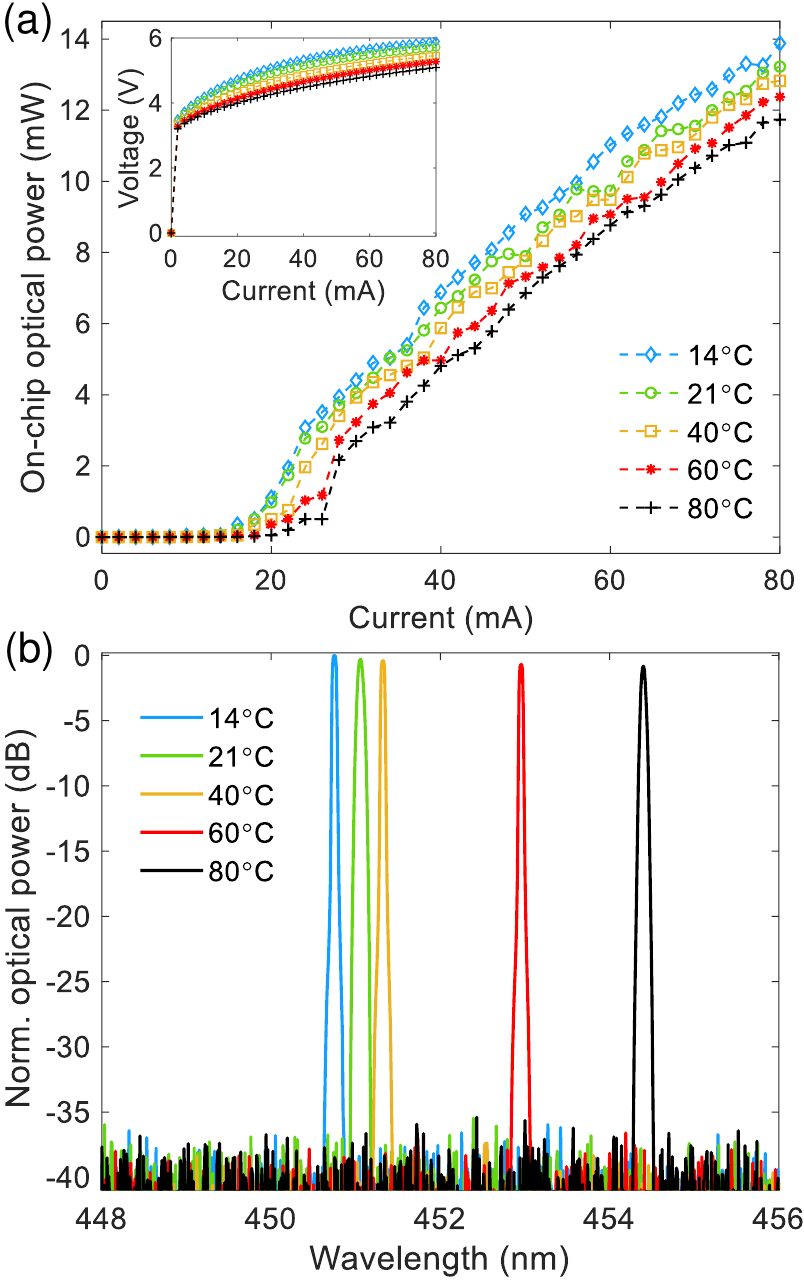}
\caption{\justifying
Temperature-dependent performance of a representative flip-chip bonded laser diode. (a) L-I curves at various temperatures. (Inset) Corresponding I-V curves. The L-I-V curves were averaged over 3 measurements. (b) Optical spectrum at temperatures corresponding to (a); 80-mA drive current, normalized to the peak power at 14 $\degree$C.
\label{fig6}}
\end{figure}

The emission power, voltage drop, and optical spectra of flip-chip bonded LDs were measured over a temperature range of 14 $\degree$C (above the dew point) to 80 $\degree$C, with the heat sink temperature adjusted by a TEC controller (Arroyo Instruments, 5240 Series). Figure \ref{fig6} shows the L-I-V curves and optical spectra of a representative LD sample (different from that in Figure \ref{fig5}). The L-I curves in Figure \ref{fig6}(a) indicate an increase in threshold current (from 16 mA to 22 mA) and a decrease in on-chip optical power ($\approx$15.5\% drop at an 80 mA drive current) as the heat sink temperature was increased. The relatively small optical power reduction at elevated temperatures with high drive currents is attributed to efficient heat sinking through the direct thermal path from the LD metal contacts to the Si substrate (via the AuSn solder and UBM). The characteristic temperature of this LD sample, calculated from the measured L-I data, was estimated to be 187 K, consistent with previously reported values for blue InGaN laser diodes with emission wavelengths between 440 nm and 450 nm \cite{bruninghoff2009temperature,wen2015enhanced}. Additionally, the inset in Figure \ref{fig6}(a) shows that the voltage across the LD decreased with increasing temperature, with a $\approx$13.6\% drop at an 80 mA drive current, aligning with the temperature-dependent voltage behavior of blue InGaN LDs reported in Refs. \citenum{ryu2006highly} and \citenum{meneghini2009analysis}.

Lastly, Figure \ref{fig6}(b) presents the temperature dependence of the flip-chip bonded LD emission spectrum. A red shift in the peak wavelength was observed as the heat sink temperature increased. The spectral linewidth remained between 0.05 nm and 0.10 nm over the tested temperature range. In agreement with Figure \ref{fig6}(a), a small decrease in peak optical power was observed at higher temperatures. 

\subsection{Performance summary and comparison}
\label{performance comparison}

The flip-chip bonded, single-spatial-mode, 450-nm InGaN laser diodes achieved high waveguide-coupled on-chip optical powers of up to 60.7 mW and on-chip wall-plug efficiencies up to 7.8\%. The series resistance and emission power of the LDs showed only minor changes after flip-chip assembly. Together, these results demonstrate the high peak performance achievable with this technique while indicating minimal damage to the LDs during hybrid integration. The primary limitation of this demonstration is the variability in post-bonding alignment and thus on-chip optical power, which constrains the repeatability and assembly yield of the current passive-alignment flip-chip bonding approach. Improvements in the assembly yield are expected using die bonders with higher placement accuracy and laser-to-waveguide coupler designs with larger misalignment tolerance. Based on the simulated alignment sensitivities in Figure \ref{fig3}, improving placement precisions from the standard deviations reported in Figure \ref{fig4} (635 nm for X-axis and 665 nm for Y-axis) to 300 nm (the placement accuracy of die bonder used in Ref. \citenum{marinins2022wafer}) would reduce the excess coupling loss by 0.16 dB (X-axis) and 2.1 dB (Y-axis) for inverse taper couplers, and by 0.17 dB (X-axis) and 1.3 dB (Y-axis) for taper couplers. Additionally, multi-tip edge couplers have been reported for the visible spectrum, with simulated 1-dB in-plane misalignment tolerances of 2.6 $\upmu$m (lateral, Y-axis) and 3.5 $\upmu$m (longitudinal, X-axis) \cite{kluge2022flip}, offering substantial improvements over the tolerances listed in Table \ref{tab:Table1}. 

Table \ref{tab:Table4} compares this work with other reported hybrid laser integration demonstrations using passive-alignment flip-chip bonding at both visible and near-infrared wavelengths. Despite higher laser-to-waveguide alignment sensitivity at shorter wavelengths (where optical mode dimensions decrease) in addition to the limited placement accuracy of our die bonder, this work presents on-chip optical powers of up to 60.7 mW and a minimum coupling loss of 1.1 dB with flip-chip bonded blue LDs. These results are comparable to the best-reported performance achieved with passive-alignment flip-chip bonding technology at near-infrared wavelengths \cite{theurer2019flip}. In comparison to similar demonstrations in the visible spectrum \cite{kluge2022flip,geuzebroek2025small}, this work achieves $\gtrsim$6$\times$ higher on-chip optical power.   

\begin{table*}
\caption{\justifying Comparison of hybrid laser/amplifier integration demonstrations using passive-alignment flip-chip bonding. InP: indium phosphide, DFB: distributed feedback, GaAs: gallium arsenide, SOA: semiconductor optical amplifier,  RGB: red, green and blue.}
\centering
\begin{tabular}{m{4.5em} m{6.5em} m{5em} m{6.5em} m{6.3em} m{14em} m{6em}}
Reference 	& LD/SOA type 	& Photonic waveguide & Wavelength (nm)& Placement accuracy & Max. CW on-chip optical power  & Min. coupling loss \\ 
\midrule
\cite{moscoso2017hybrid} & 
LD  & Si	&	Near infrared & Sub-micron & $\approx$3 mW & 7 dB \\[5pt]
\cite{theurer2019flip} & InP DFB LD & SiN  & $\approx$1550 & Sub-micron & $\approx$40 mW & 3 dB$^a$ 	\\[5pt]
\cite{marinins2022wafer} & InP DFB LD & SiN & $\approx$1550 & Sub-300 nm & 40 mW (with epoxy underfill)& 1.1 dB \\[5pt]
\cite{kluge2022flip} & GaAs SOA & SiN & $\approx$640 & Sub-2 $\upmu$m\ & Lasing operation not achieved & 7.1 dB \\[5pt]
\cite{geuzebroek2025small} & RGB LDs & SiN & RGB & Sub-500 nm & $\approx$10 mW$^b$ & N.A.\\[10pt]
This work & InGaN LD & SiN & $\approx$450 & Sub-micron & 60.7 mW & 1.1 dB \\
\end{tabular}
\justifying
\footnotesize{
$^a$Results for one sample reported. 
\\
$^b$Free-space emission from the integrated photonic chip, combining 3 LDs (R, G, B).}
\label{tab:Table4}
\end{table*}

\section{Photonic integrated circuit demonstration} 

\begin{figure*}[h]
\includegraphics[width=\textwidth]{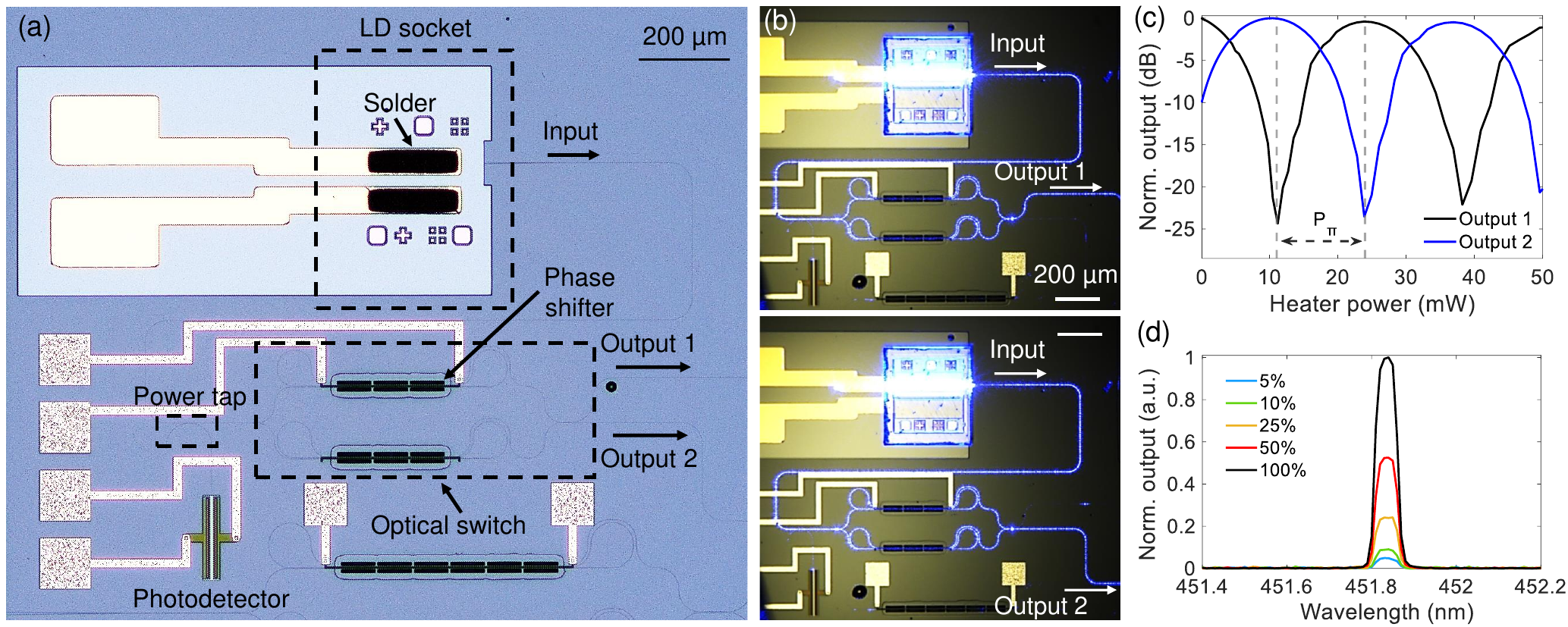}
\caption{\justifying
Photonic integrated circuit (PIC) with a flip-chip bonded LD, an integrated photodetector, and a thermo-optic (TO) switch. (a) Micrograph of the PIC prior to laser bonding. (b) Micrographs of the PIC (after laser bonding) showing operation of the bonded laser and switching between the output waveguides (Outputs 1 and 2). (c) TO switching operation: normalized output optical power from the PIC vs. heater power dissipation of the TO phase shifter. (d) Variable optical attenuation using the switch: normalized optical spectra of PIC Output 1 at various optical power levels. The laser drive current was 35 mA.
\label{fig7}}
\end{figure*}

In this section, we present a proof-of-concept active photonic integrated circuit combining our Si photonic platform’s thermo-optic switching and photodetection capabilities with a hybrid-integrated InGaN LD. An LD was flip-chip bonded into a socket on the PIC following the method in Section \ref{process}. Emitted light was routed through a SiN power tap that diverted a small fraction to a waveguide-coupled photodetector for power monitoring, followed by a thermo-optic switch, Figure \ref{fig7}(a). The power tap was implemented using an evanescent directional coupler with a designed power coupling of $\approx$1\%. A 200-nm-wide SiN waveguide was routed underneath a 200-$\upmu$m-long Si PIN junction to form the photodetector \cite{govdeli2024broadband}. The optical switch used a Mach-Zehnder interferometer configuration with a 324-$\upmu$m-long TO phase shifter in each arm. The TO phase shifter featured a TiN heater along with a 3-pass folded waveguide and thermal isolation trenches to enhance phase modulation efficiency, Figure \ref{fig1}(a); no undercut etching was performed. The switch supported optical switching between two output ports (Outputs 1 and 2), as well as variable optical attenuation at one selected output port. The output ports were routed to edge couplers at a chip facet for coupling to a single-mode optical fiber. Tungsten needle probes contacted the UBM and aluminum pads on the Si photonic chip to apply electrical drive signals.

\subsection{Optical switching and variable attenuation}
\label{switching}

Figure \ref{fig7}(b) demonstrates the optical switching operation of the PIC, achieved by driving the LD and TO phase shifter. The normalized PIC optical output power as a function of heater power dissipation is plotted in Figure \ref{fig7}(c). The extinction ratio (ER) of the optical switch and heater power for a $\pi$-radian phase shift (P$_\pi$) in the phase shifter were calculated from the first half-period of Figure \ref{fig7}(c). Outputs 1 and 2 exhibited ERs of 24.4 and 23.5 dB, with corresponding P$_\pi$ values of 12.7 and 13.4 mW, respectively. Although the switching speed was not characterized in this work, a similar TO switch has been reported to exhibit a switching speed of 11.8 kHz \cite{alemany2021thermo}.

The optical switch also functions as a variable optical attenuator (VOA) when a single output port is used. This functionality is valuable in applications requiring precise, high-dynamic-range optical power control without associated wavelength shifts, such as laser-based displays. Such control is difficult to achieve through LD current tuning because of the dependence of wavelength on current and the nonlinearity of the L–I characteristics near threshold, Figure \ref{fig5}. To demonstrate VOA operation with the PIC, the LD was driven at a constant current of 35 mA, while the heater power applied to the phase shifter was varied to control the optical power at Output 1. Figure \ref{fig7}(d) shows the normalized optical spectra from Output 1 at different power levels relative to the peak optical power. The peak emission wavelength ($\approx$451.8 nm) and spectral linewidth (0.05 - 0.10 nm) remained stable across all measured optical power levels (spanning 13 dB of attenuation), owing to the constant LD drive current.

\subsection{On-chip power monitoring} 
\label{power monitoring}

\begin{figure*}[t]
\centering
\includegraphics[width=0.8\textwidth]{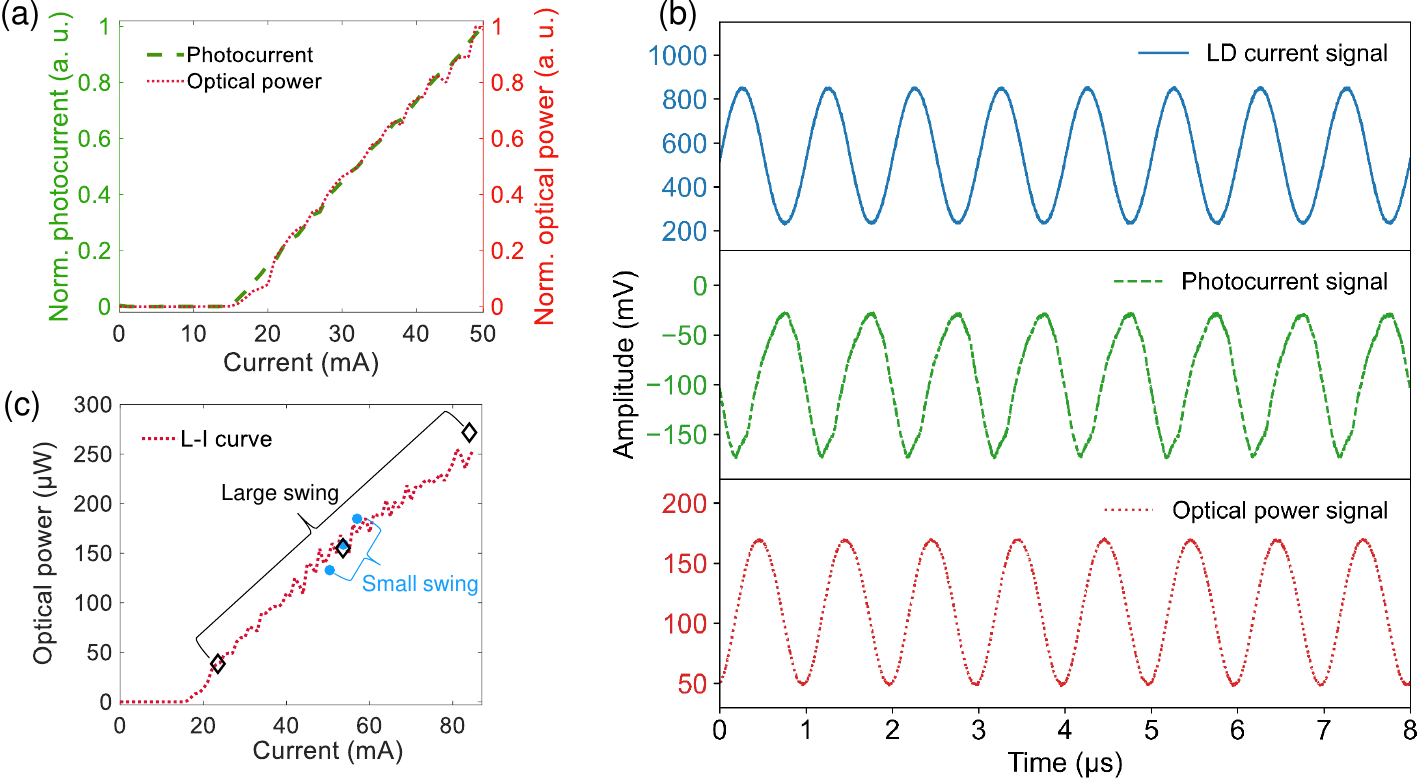}
\caption{\justifying
PIC power monitoring demonstration using the integrated photodetector in Figure \ref{fig7}(a). (a) Normalized DC photocurrent and output optical power from the PIC vs. LD drive current; data averaged over 7 measurements. (b) Comparison of signals acquired by the oscilloscope during sinusoidal LD current modulation at 1 MHz. (c) Comparison of off-chip optical power with DC drive current (L-I curve) and dynamic optical power during LD modulation (maximum, average, and minimum of the sinusoidal modulation). The L-I curve was averaged over 4 measurements. Small- and large-swing LD modulation cases are shown, distinguished by colored brackets. 
\label{fig8}}
\end{figure*}

The active functionality of the PIC was further tested through on-chip power monitoring of the flip-chip-bonded LD emission using the integrated photodetector. The photocurrent from the on-chip photodetector was characterized under both DC and AC LD driving conditions. DC power-monitoring characteristics are shown in Figure \ref{fig8}(a), comparing on-chip photocurrent with off-chip power measurements as the LD drive current was varied. The traces are normalized to facilitate comparison of their shapes. The photocurrent and optical power curves are consistent, both exhibiting high linearity above the lasing threshold.  

For AC modulation, the on-chip photocurrent was compared with the LD drive current waveform and with a reference off-chip optical power measurement recorded using an external detector. In these measurements, the LD was DC-biased via a source measure unit and modulated with a 1 MHz sine-wave AC signal from an arbitrary function generator (AFG, Tektronix AFG31000). The DC bias and AC modulation signals were combined using a bias tee (Pasternack, PE1BT1002) to drive the LD, with a 10-$\Omega$ sensing resistor connected in series to measure the current during modulation. The integrated photodetector was reverse-biased at 2 V using a second channel of the source measure unit, and its photocurrent was amplified by a transimpedance amplifier (TIA, Koheron TIA100-2k-SMA). The modulated optical output of the PIC was collected through a single-mode fiber and detected by an off-chip Si switchable-gain detector (Thorlabs, PDA36A2). The LD current (measured via the sensing resistor in series), photocurrent from the integrated photodetector, and off-chip optical power were recorded simultaneously with an oscilloscope (Keysight, DSOX4054A). A schematic of the measurement setup is shown in Figure \ref{figS7} in the Supplementary Materials.

During modulation, the LD was biased at a DC current of 53 mA to ensure stable emission above the lasing threshold while avoiding thermal rollover within the modulation swing. Both small and large current swings were tested, with AFG voltage amplitudes of 0.5 V$_{pp}$ and 4.5 V$_{pp}$, respectively. The measured LD current, on-chip photocurrent, and off-chip optical power signals during LD modulation for the large current swing case are shown in Figure \ref{fig8}(b). The photocurrent signal was synchronized with the LD current signal, exhibiting a phase inversion attributed to the inverting operation of the TIA. The off-chip optical power is shown in the bottom panel of Figure \ref{fig8}(b), with a slight phase shift relative to the LD current signal, likely due to the internal phase delay of the switchable-gain detector. All three signals maintained a consistent frequency of 1 MHz with sinusoidal waveforms.

Calculations and calibration measurements were used to determine the amplitudes of the LD current and off-chip optical power during modulation from the oscilloscope voltage traces. Using the known sensing resistor value, the LD current was found to vary from $\approx$23.5 mA to $\approx$84.0 mA for the large current swing and from $\approx$50.4 mA to $\approx$57.0 mA for the small current swing. Next, the procedure described in Section \ref{calibration} of the Supplementary Materials was applied to extract the magnitude of the optical power modulation measured by the off-chip detector (referred to as the dynamic optical power). Figure \ref{fig8}(c) compares the DC off-chip optical power vs. current (static L–I curve averaged over 4 measurements) with the maximum, average, and minimum off-chip optical powers during AC modulation. Both the small and large current swing cases are shown in Figure \ref{fig8}(c). The derived dynamic power values closely follow the static L–I curve and indicate optical power modulations (ratio of maximum to minimum) of $\approx$7.1$\times$ and $\approx$1.4$\times$ for the large and small current swings, respectively. Small differences between static and dynamic optical powers arise from the inherent LD temperature variations in these operating conditions.

\subsection{Integration outlook}

The demonstrated integration of visible-light LDs with on-chip thermo-optic tuning and photodetection provides a practical and fundamental set of functionalities for visible-spectrum PICs, particularly for enabling wavelength-tunable lasers, which have diverse applications such as quantum information processing \cite{newman2019architecture}, spectroscopy \cite{li2018high}, and biosensing \cite{guo2020hyperboloid}. Recent work on visible-light LDs and gain chips actively-aligned to SiN PICs have demonstrated wavelength-tunability via thermo-optically controlled feedback \cite{franken2021hybrid,corato2023widely}. The integration of photodetectors, as demonstrated here, opens new opportunities for fully-integrated wavelength-tunable lasers. In principle, these on-chip devices would allow stabilization and tracking of both emission power (through power taps) and wavelength (through integrated wavemeters). Additionally, the building blocks integrated in this work are directly applicable to emerging PIC-based microdisplays for augmented and virtual reality \cite{shi2025flat,geuzebroek2025small}. In such systems, on-chip switches and VOAs could enable compact, dynamic routing to different emitters and brightness control, while integrated photodetectors could facilitate brightness monitoring, which is particularly important for ensuring eye safety. In the above applications, passive-alignment flip-chip LD bonding, combined with foundry-based Si photonics fabrication, offers a direct path to scaling PIC complexity, wafer fabrication volumes, and assembly throughput. 

As shown in our recent work \cite{alperen2025}, similar integrated photodetectors on this platform achieve up to 18 GHz bandwidth and support avalanche gain. The 1 MHz AC modulation tested in Section \ref{power monitoring} was limited by the bandwidths of the TIA (2 MHz) and the switchable-gain detector used for off-chip power measurements (1.6 MHz at 10 dB gain), not by the device itself. Future implementations of PICs incorporating arrays of directly-modulated, hybrid-integrated, visible-spectrum LDs, together with high-bandwidth, sensitive photodetectors, could form the basis of new microsystems for short-reach interconnects \cite{pezeshki202510gbps} and underwater communications \cite{notaros2023liquid}. 

Though not demonstrated here, our visible-light Si photonic platform also supports microelectromechanical systems (MEMS) devices for beam scanning. These can be realized either by undercut etching the Si substrate to release movable SiO$_2$-cladding cantilevers, or by wafer thinning to define larger structures with a robust Si backbone. Integrating waveguide emitters into these devices enables compact beam scanners; we have recently demonstrated both electrothermally actuated SiO$_2$ scanners \cite{sharif2023microcantilever} and Si MEMS actuated by co-packaged piezoelectric \cite{ankita2025}. Integration of these devices with on-chip lasers would enable compact, fully integrated beam-scanning systems, opening new possibilities for applications such as next-generation microdisplays and advanced scanning microscopy systems \cite{mashayekh2025silicon}.

\section{Conclusion} 

In summary, we have demonstrated the hybrid integration of blue, 450-nm-wavelength, InGaN laser diodes onto a foundry-fabricated visible-light Si photonics platform using passive-alignment flip-chip bonding. Leveraging the co-design of the laser diodes and Si photonic chips, which included precisely patterned alignment marks and mechanical stoppers, the flip-chip bonding process achieved sub-micron in-plane laser-to-chip alignment accuracy. Across the 40 flip-chip-bonded laser samples, the measured optical power coupled to the Si photonic chip reached 60.7 mW. A highest on-chip wall-plug efficiency of 7.8\% and a minimum coupling loss of 1.1 dB were achieved. Thermal characterization showed stable operation of the flip-chip bonded lasers at elevated temperatures up to 80 $\degree$C, maintaining high optical output power. A proof-of-concept active photonic integrated circuit was demonstrated, incorporating a flip-chip bonded laser diode, an integrated waveguide-coupled Si photodetector as an on-chip power monitor, and a thermo-optic switch for dynamic optical routing and variable optical attenuation. 

Overall, this work establishes a new benchmark for scalable hybrid integration of visible-wavelength laser diodes on Si photonic platforms. Although challenges remain in reducing variability in laser-to-chip alignment, advances in high-precision die bonding and laser-to-waveguide coupler designs present clear pathways for further improvement. The localized heating used in the flip-chip bonding method is compatible with integrating multiple laser diodes on a single PIC, and future efforts will target incorporating lasers at multiple wavelengths. We anticipate that the combined functionalities of our platform, including hybrid-integrated lasers, thermo-optic devices, photodetectors, and MEMS devices, will enable fully-integrated wavelength-tunable laser arrays, as well as complex PICs for applications in augmented and virtual reality, quantum information processing, biophotonics, and data communications.

\begin{acknowledgement}
The authors thank Blaine J. McLaughlin at the Max Planck Institute of Microstructure Physics for polarization extinction ratio measurements.
\end{acknowledgement}

\begin{funding}
This work was supported by the Max Planck Society. The authors R. Lawrowski and M. Jama gratefully
acknowledge partial financial support by the German Federal Ministry of Education and Research (BMBF) in the
funding program Trusted Electronics (ZEUS) within the joint project VE-Silhouette under grant ME1ZEUS006.
\end{funding}

\begin{authorcontributions}
MJ and WDS designed and organized the project. JCM and WDS designed the Si photonic chips. HC and GQL were responsible for fabrication of the Si photonic wafers. RL and MJ designed and fabricated the InGaN laser diodes. XM simulated the laser-to-waveguide coupling and analyzed the misalignment tolerance. XM and FW developed the flip-chip bonding process and performed the integration procedure. XM and HW measured the post-bonding misalignment. XM and JNS built the laser characterization setup. XM tested the flip-chip bonded lasers diodes and photonic integrated circuit. XM and PK tested epoxy underfill and encapsulation of the laser diodes. GQL, JKSP, MJ, WDS supervised the project. XM and WDS prepared the manuscript with support from all authors. All authors have accepted responsibility for the entire content of this manuscript and approved its submission.
\end{authorcontributions}

\begin{conflictofinterest}
Authors state no conflict of interest.
\end{conflictofinterest}

\begin{dataavailabilitystatement}
The datasets generated during and/or analyzed during the current study are available from the corresponding authors on reasonable request.
\end{dataavailabilitystatement}

\onecolumn
\newpage
\section*{Supplementary materials} 
\label{supplementary}

\setcounter{figure}{0}
\setcounter{table}{0}

\renewcommand {\thesubsection}
{S\arabic{subsection}}

\subsection{Rotational misalignment of flip-chip bonded laser diodes}

\begin{figure*}[!htbp]
\centering
\renewcommand{\thefigure}{S\arabic{figure}}
\includegraphics[width=0.8\textwidth]{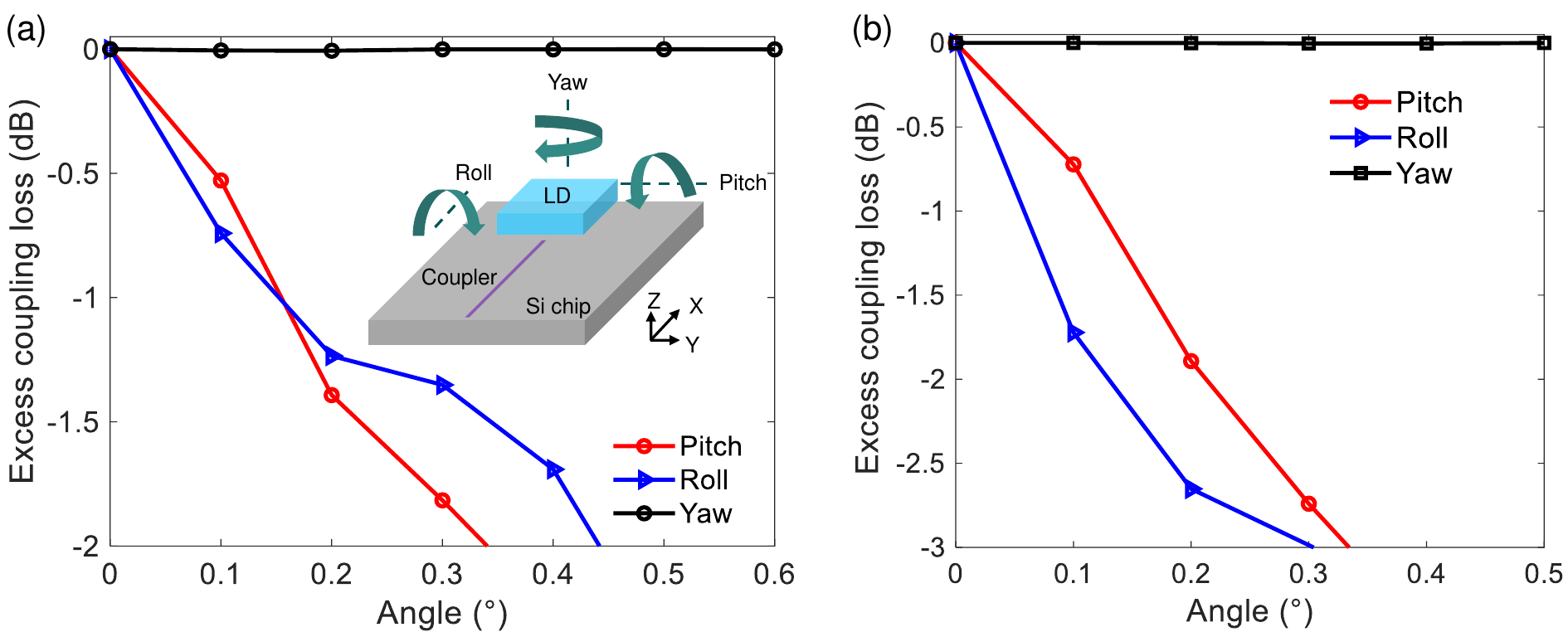}
\caption{\justifying
Simulation of rotational misalignment between the laser diodes (LDs) and silicon nitride (SiN) waveguide couplers for (a) inverse taper and (b) taper coupler designs. Inset: illustration of the rotation axes.
\label{figS1}}
\end{figure*}

To investigate the influence of rotational misalignment between the laser diode (LD) and the silicon nitride (SiN) waveguide coupler, we simulated the excess coupling loss caused by angular tilts along the pitch, roll, and yaw axes, as shown in the inset of Figure \ref{figS1}(a). The coupling loss was calculated with three-dimensional finite-difference time-domain (FDTD) electromagnetic simulations (Ansys Lumerical). The simulations indicate that taper couplers in Figure \ref{figS1}(b) are more sensitive to rotational misalignment compared to inverse taper couplers in Figure \ref{figS1}(a). As detailed in Section \ref{bonding performance}, among the 40 flip-chip bonded LD samples, the measured angular tilts were minimal, corresponding to simulated excess coupling losses of $<$0.5 dB along all three rotation axes for both coupler designs.

\begin{figure*}[!htbp]
\centering
\renewcommand{\thefigure}{S\arabic{figure}}
\includegraphics[width=1\textwidth]{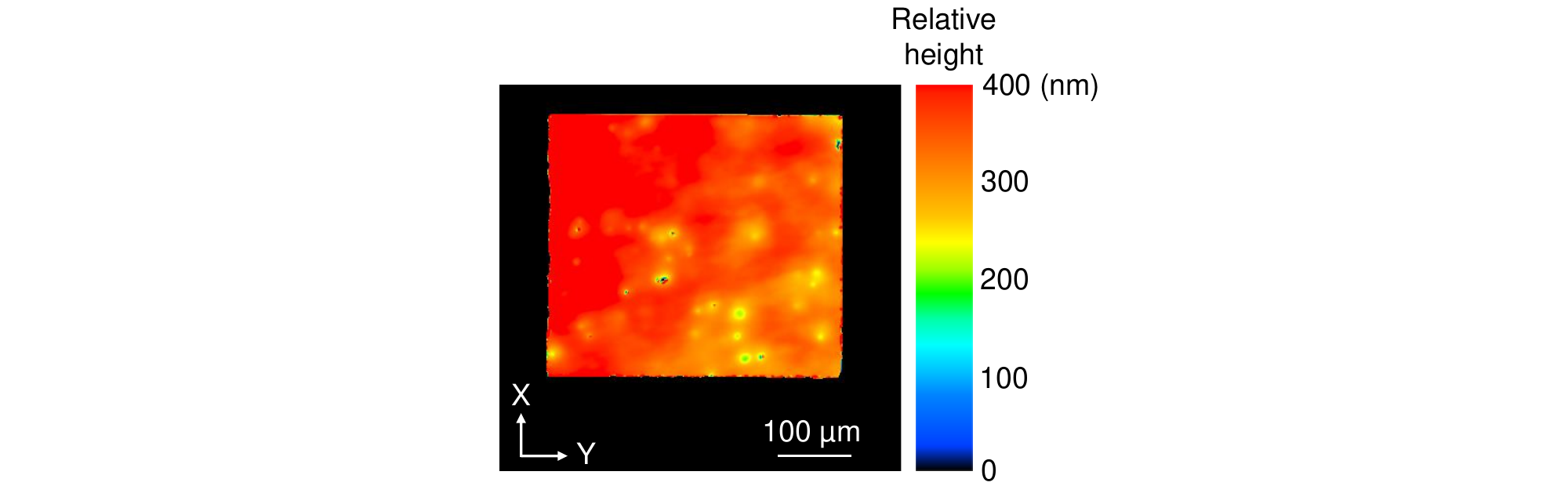}
\caption{\justifying
Height profile measurement of a flip-chip bonded laser diode using white light interferometry. The height offset between the reference plane on the silicon (Si) photonic chip (shown in black) and the top surface of the laser diode (colored) was subtracted to show the relative height difference. The displayed height range was adjusted to improve visibility of the measured relative height difference.
\label{figS2}}
\end{figure*}

We also evaluated the rotational misalignment [pitch and roll axes; yaw-axis tilt calculated based on the 4 alignment marks pairs as shown in Figure \ref{fig4}(a)] by measuring the height of flip-chip bonded laser diodes. Figure \ref{figS2} shows the top surface profile of a representative flip-chip bonded laser diode. The height was measured with white light interferometry using a 3D surface profiler (Keyence, VK-X3000). The vacant area near the measured LD bonding socket on the silicon (Si) chip served as the reference plane. The upper and lower limit of the displayed heights were adjusted for better visibility of the height variation on the LD, which was small compared to the $\approx$100 $\upmu$m LD thickness. In addition, the height offset between the reference plane on the Si chip and the top surface of laser diode was subtracted during the measurements, showing the relative height difference. The maximum height difference along the X  and Y axes of the LD shown in Figure \ref{figS2} are $\approx$130 nm and $\approx$180 nm, respectively, corresponding to angular tilts of $<$0.02$\degree$ (pitch) and $<$0.03$\degree$ (roll). The yaw-axis tilt was measured as 0.04$\degree$ for this sample.

\subsection{Bonding of multiple LDs on a single Si photonic chip}
\label{multi-die}

Localized heating from the pickup tool in the flip-chip bonding process enables independent bonding of multiple LDs onto a single Si photonic chip. However, when bonding sockets are in close proximity, thermal exposure from successive bonding steps may influence the alignment of previously bonded LDs. To assess this potential impact, additional data analysis is presented below.

The Si photonic chips used to test flip-chip bonding of LDs (Sections \ref{bonding method} and \ref{LD characterization}) included multiple bonding sockets, Figure \ref{fig2}(d). Each chip contained two rows of sockets: one designed with inverse tapers and the other with tapered laser-to-waveguide couplers. The intra-row pitch was 640 $\upmu$m, while the inter-row pitch was larger at 2.8 mm. Among the 40 bonded LD samples shown in Figure \ref{fig4}(b), 26 were associated with Si photonic chips onto which 2 – 3 LDs had been bonded in the same row, whereas in the remaining cases only a single LD was bonded per Si chip. Characterization of the flip-chip bonded LDs occurred after all LDs had been bonded to the corresponding Si chip.

\begin{figure}[H]
\centering
\renewcommand{\thefigure}{S\arabic{figure}}
\includegraphics[width=0.99\textwidth]{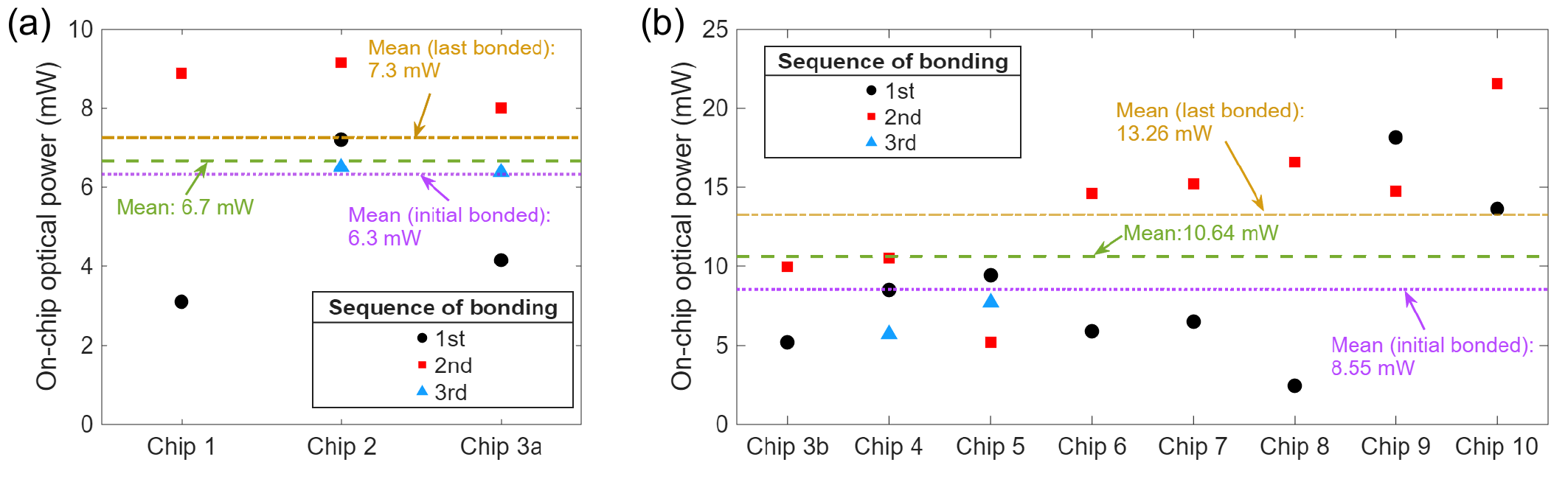}
\caption{\justifying
On-chip optical power of multiple LDs bonded on the same Si chip using (a) inverse taper and (b) taper couplers. The data correspond to a subset of samples from Figure \ref{fig4}(b). LD drive current: 50 mA. Legends denote the bonding sequence of LDs on each chip; measurements were taken after all LDs were bonded. Green dashed lines: mean on-chip optical power across all LDs per panel. Purple dashed lines: mean power of initial bonded samples (1st LD for sequences of two samples, 1st and 2nd for sequences of 3 LDs). Yellow dashed lines: mean power of last bonded samples (2nd or 3rd LD).
\label{figS3}}
\end{figure}

Evaluation of the measured on-chip optical powers with respect to the bonding sequence provides insight into the dependence of integrated LD performance on subsequent bonding steps. Figure \ref{figS3} shows the on-chip optical powers of these samples, which represent a subset of the data presented in Figure \ref{fig4}(b) (measured at an LD drive current of 50 mA). The data are categorized by Si photonic chips with LDs bonded to sockets incorporating inverse taper couplers (Chips 1–2) and tapered laser-to-waveguide couplers (Chips 4–10). On Chip 3, LDs were bonded to both rows of sockets; the corresponding data are labeled as Chip 3a and Chip 3b.

Figure \ref{figS3} compares the on-chip optical power of LDs as a function of bonding order, where the first-bonded LDs in each sequence were subsequently exposed to thermal cycles from later bonding steps. For inverse taper couplers, the mean power of initially bonded LDs was 86\% of that of last-bonded LDs. For taper couplers, the first-bonded LDs exhibited a mean power only 64\% of that of last-bonded LDs. Additionally, for the complementary set of samples (not shown in Figure \ref{figS3}) in which only a single LD was bonded per Si photonic chip, the mean on-chip optical powers ($\pm$ standard deviation) were 8.1 $\pm$ 2.6 mW for inverse taper couplers and 9.0 $\pm$ 4.7 mW for taper couplers. With respect to Figure \ref{figS3}, these values align with the mean of last-bonded LDs for inverse taper couplers and, conversely, with the mean of first-bonded LDs for taper couplers, highlighting the variability inherent to the bonding process.

While the data in Figure \ref{figS3} point to a potential influence of bonding sequence on LD performance, the observed differences are comparable to the variability seen in chips with only a single bonded LD. Refinements to the flip-chip process and coupler design, aimed at reducing coupling variability, may be necessary to determine whether such effects are significant. A larger bonding socket pitch may also reduce thermal interactions between successive bonding steps.

\subsection{Measurement of laser samples exhibiting thermal rollover}
\label{more LD_rollover}

\begin{figure}[H]
\centering
\renewcommand{\thefigure}{S\arabic{figure}}
\includegraphics[width=0.8\textwidth]{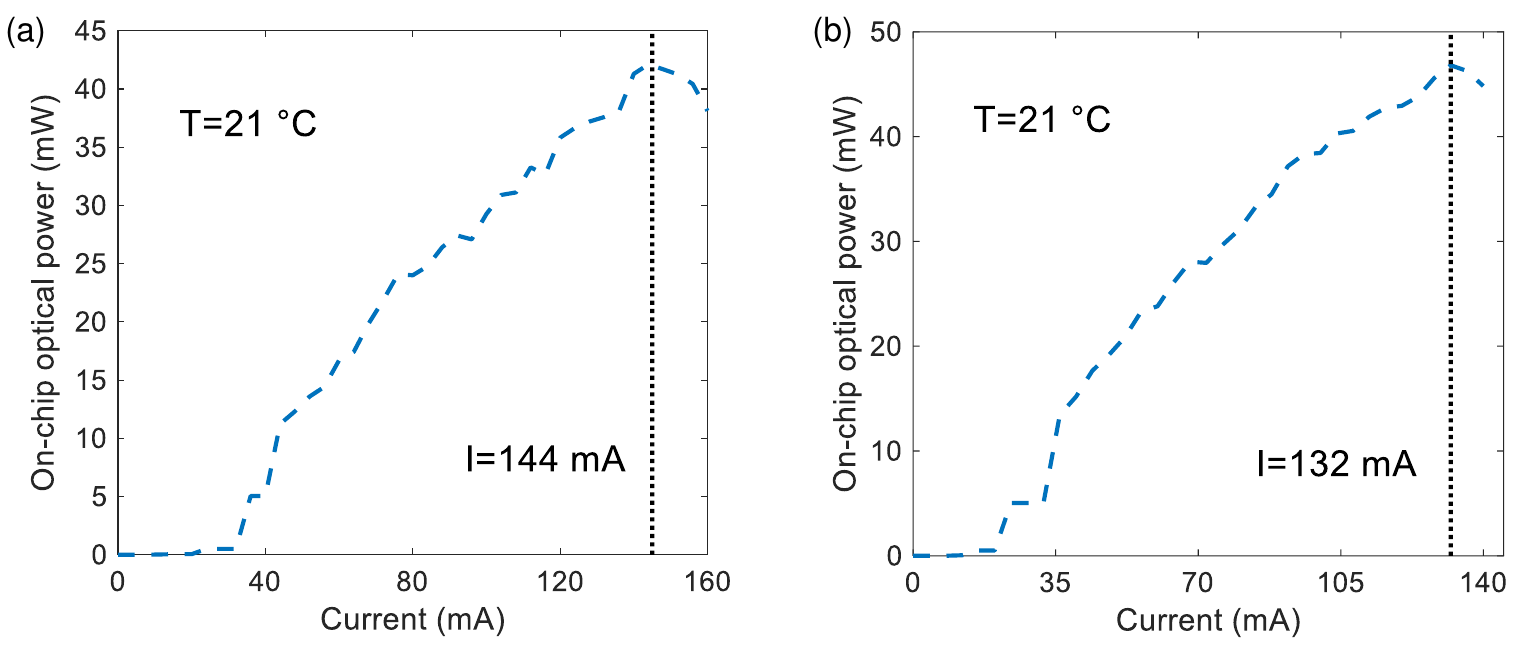}
\caption{\justifying
On-chip optical power (L) vs. current (I) curves of two representative flip-chip bonded laser diodes (LD1 and LD2). Thermal rollover was observed when the LD current exceeded 144 mA and 132 mA for (a) LD1 and (b) LD2, respectively.
\label{figS4}}
\end{figure}

The on-chip optical power vs. current (L-I) curves of two additional representative flip-chip bonded LD samples (LD1 and LD2) are shown in Figure \ref{figS4}(a) and Figure \ref{figS4}(b), respectively. Both samples used SiN taper designs for the LD-to-waveguide couplers. Thermal rollover was observed when the drive current exceeded 144 mA for LD1, with a peak on-chip optical power of 42.1 mW and corresponding on-chip wall-plug efficiency of $\approx$5.5\%. LD2 exhibited thermal rollover at drive currents $>$132 mA, with a peak on-chip optical power of 46.8 mW and on-chip wall-plug efficiency of $\approx$7.8\%. For both LD samples, the L-I curves were reproducible (under low drive currents without thermal rollover) before and after the onset of thermal rollover, indicating stable LD operation without observable LD degradation.

\subsection{Epoxy underfill and encapsulation of bonded lasers}
\label{epoxy}

\begin{figure*}[!htbp]
\centering
\renewcommand{\thefigure}{S\arabic{figure}}
\includegraphics[width=0.85\textwidth]{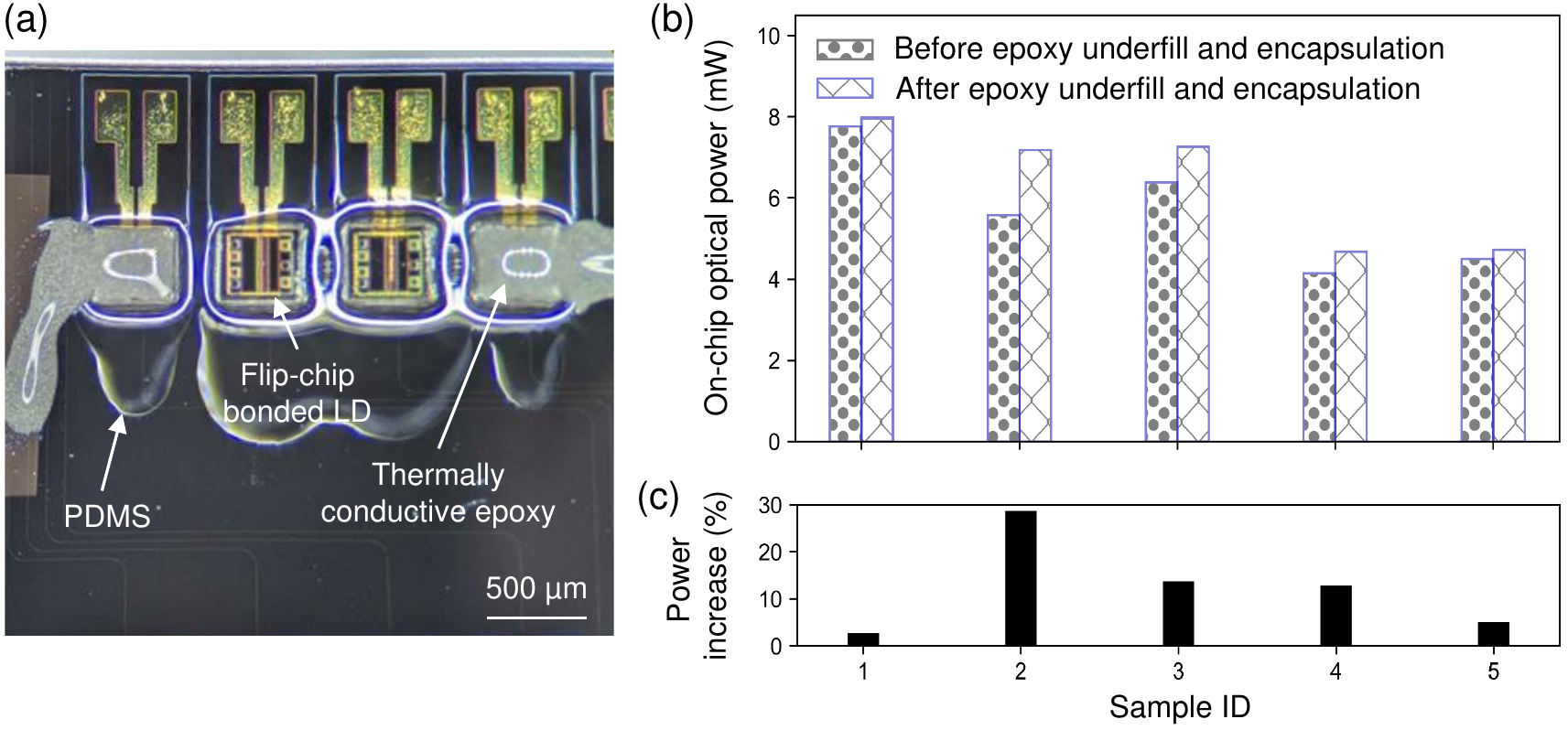}
\caption{\justifying
(a) Micrograph of flip-chip bonded LDs with underfill (with PDMS) and encapsulation (with thermally conductive epoxy). The two center LDs were underfilled, while the outer LDs underwent both underfill and encapsulation. (b) On-chip optical power comparison (at 50 mA LD current) before and after epoxy underfill and encapsulation. (c) Percent increase in optical power for the five LD samples after epoxy underfill and encapsulation.
\label{figS5}}
\end{figure*}

We investigated a post-bonding epoxy underfill and encapsulation process to enhance the on-chip optical power of flip-chip-bonded LDs while improving mechanical robustness. Epoxy underfill was applied to five flip-chip-bonded LD samples to improve optical coupling to the SiN waveguide by reducing the large refractive index discontinuity at the LD–waveguide air gap. An optically transparent adhesive was first applied around the flip-chip-bonded LD using a thin metal rod applicator mounted on a micro-manipulator for precise control of the placement and volume. The bonded LDs were then encapsulated with thermally conductive epoxy to improve heat dissipation. Encapsulated devices are shown in Figure \ref{figS5}(a). 

Various optically clear adhesives were evaluated for the underfill step. The UV-curable epoxies Delo OB6268 and Dymax 9622 caused loss of light transmission in the SiN waveguides when the LD current exceeded 45 mA (LD emission power 20 – 25 mW). This effect is likely due to degradation or burning of the adhesive under high-intensity blue light. Similar observations have been reported in our previous work \cite{roszko2025foundry}. To avoid this limitation, thermally-curable polydimethylsiloxane (PDMS, Sylgard 184) mixed at a 10:1 ratio (pre-polymer to cross-linker ratio) was chosen as the underfill epoxy. PDMS was thermally cured at 60 $\degree$C for 1 hour after application. With PDMS filling the LD-waveguide gap, the calculated Fresnel reflection at the coupling interface is estimated to decrease from $\approx$24.8\% to $\approx$9.4\% \cite{li2015improved,cai2013new}.

After underfill with PDMS, the electrically insulating, thermally conductive epoxy, Duralco 128, was then applied over and around the LD to improve heat dissipation, followed by curing at 60 $\degree$C for 1 hour. A low curing temperature of 60 $\degree$C was chosen to avoid thermal damage to the LD while maintaining a moderate curing time. Both PDMS and Duralco 128 can also be cured at room temperature with an extended curing time. Different thermally conductive epoxies were evaluated following the PDMS underfill step. Among the tested epoxies (Duralco 128, Duralco 132, and Loctite Ablestik 2151), Duralco 128 showed the largest increase in optical output power, likely due to better thermal dissipation. The epoxy shown in Figure \ref{figS5}(a) is Duralco 132. 

On-chip optical powers before and after PDMS underfill and Duralco 128 encapsulation are summarized in Figure \ref{figS5}(b). As shown in Figure \ref{figS5}(c), the on-chip optical power increased by $\approx$2.7\% to $\approx$28.7\% across the five samples after PDMS underfill and thermal epoxy encapsulation. The average power increase across the samples was 12.6\%.

\subsection{Impact of the flip-chip process on laser diode characteristics}
\label{bonding impact}

\begin{figure}[H]
\centering
\renewcommand{\thefigure}{S\arabic{figure}}
\includegraphics[width=0.55\textwidth]{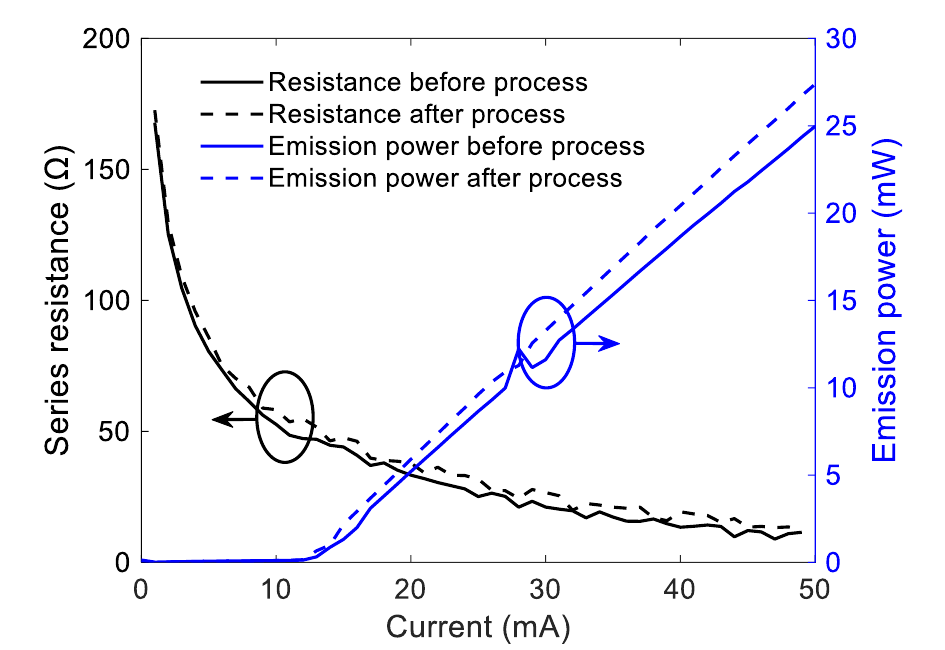}
\caption{\centering
Series resistance and emission power of a representative LD before and after heat and pressure exposure.
\label{figS6}}
\end{figure}

Additional experiments were performed to investigate the impact of heat and pressure from the die bonder on LD performance. To directly measure LD emission power, independent of alignment-sensitive waveguide coupling, the LDs were mounted face-up on a Si dummy chip without solder. This configuration limited testing to an inverted arrangement, where the pickup tool applied heat and pressure to the LD top (contact) surface rather than the bottom surface (as in the flip-chip bonding procedure). Series resistance and emission power were measured before and after exposure.

First, to measure the LD characteristics prior to heat and pressure application (exposure process), three LDs were positioned with metal contacts facing up near the facet of a dummy Si chip (identical dimensions as the Si photonic chips used in the flip-chip bonding process). A thin layer of uncured metal adhesive (Loctite, Ablebond 84-1LMIT1) secured the LDs to the Si chip. Without curing of the metal adhesive, L-I-V curves before the exposure process were collected with a source measure unit (SMU) and a power meter, with DC needles probing the P and N contacts of the LDs. The LD emissions were directly incident on a free-space detector for emission power measurements. Next, the LDs on the Si dummy chip were mounted in the die bonder, and the pickup tool touched down onto each LD, following the same heating and force profile as the flip-chip bonding process. As in that process, heat from the die bonder stage was also applied. L-I-V curves of the LDs were then measured using the same procedure as prior to the exposure process. 

Figure \ref{figS6} compares the series resistance and emission power of a representative LD sample before and after heat and pressure exposure from the die bonder, with the resistance extracted from the measured I–V curve. After the process, the series resistance increased slightly, by $\approx$14.7\% (3.8 $\ohm$) on average compared to its initial values. The emission power at 50-mA drive current increased slightly ($\approx$9.7\%) after the process, which may be attributed to thermal curing of the metal adhesive. The applied pressure during the exposure process likely pressed the LD and dummy Si chip into closer contact, which may have altered the thermal conduction properties through the cured metal adhesive. The other two LD samples showed similar results of increased series resistance ($\approx$14.4\% and $\approx$11.6\% increase on average) and emission power ($\approx$12.8\% and $\approx$17.3\% increase on average). Overall, the results imply that changes in LD performance during the flip-chip bonding process are likely to be small. A limitation of these tests is the manner in which force was applied. Here, force was exerted on the top surface of each LD while the bottom surface contacted the Si dummy chip. By contrast, during flip-chip bonding, force was applied to the back of the LD with its stoppers contacting the Si photonic chip.

\subsection{Schematic of the measurement setup for on-chip power monitoring}
\label{setup}

\begin{figure}[H]
\centering
\renewcommand{\thefigure}{S\arabic{figure}}
\includegraphics[width=0.8\textwidth]{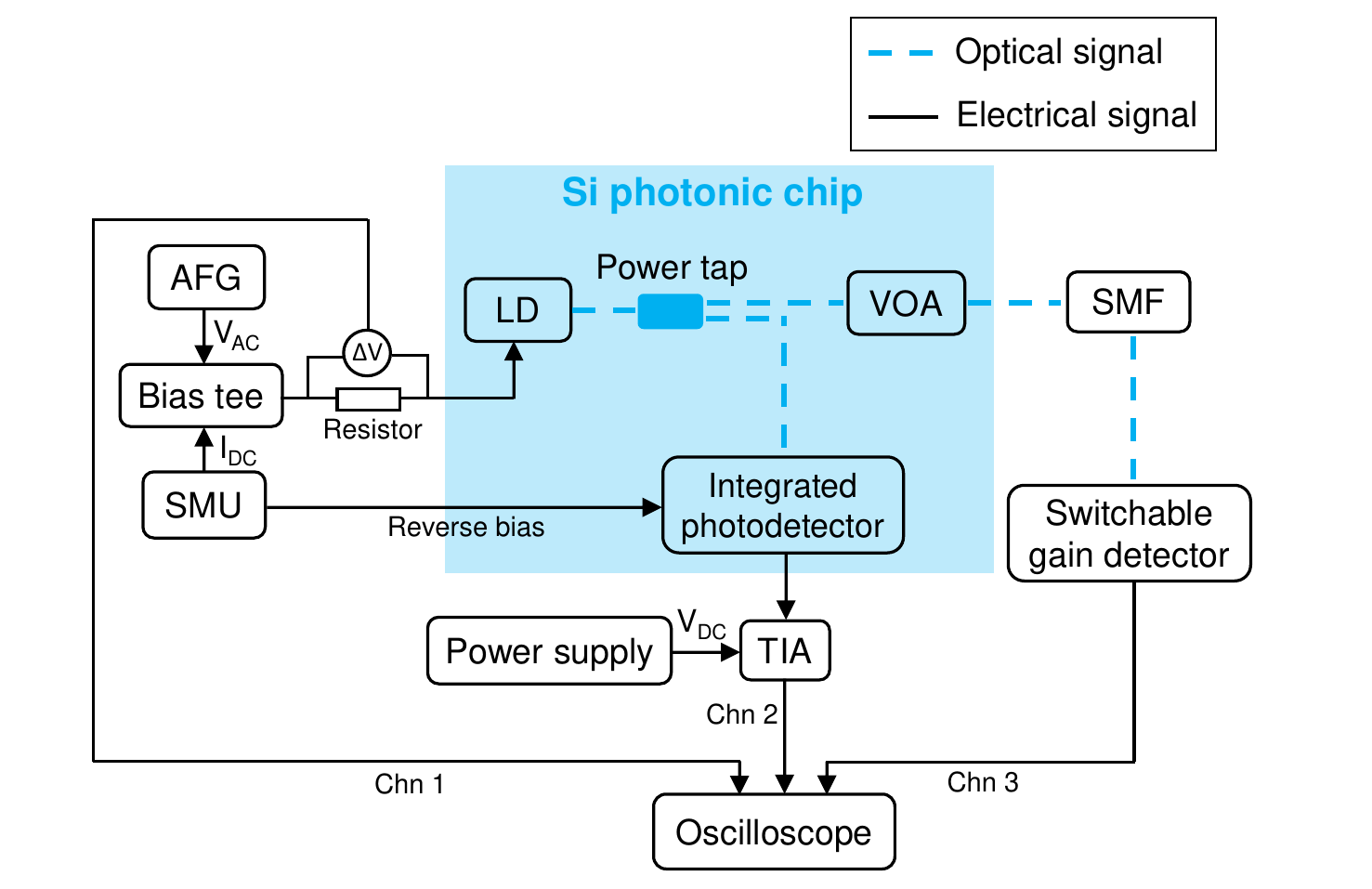}
\caption{\justifying
Schematic of the measurement setup for power monitoring with the photonic integrated circuit (Figure \ref{fig8}). AFG: arbitrary function generator, SMU: source measure unit, VOA: variable optical attenuator, TIA: transimpedance amplifier, SMF: single-mode fiber, Chn: oscilloscope channel.
\label{figS7}}
\end{figure}

Figure \ref{figS7} illustrates the experimental setup used for the on-chip power monitoring demonstration in Figure \ref{fig8}. The photonic integrated circuit (PIC, on the Si photonic chip) consisted of a flip-chip bonded LD, an optical power tap, an integrated photodetector used as a power monitor, and a thermo-optic switch used as a variable optical attenuator (VOA). A two-channel source measure unit delivered a DC current bias to the LD and a reverse bias voltage to the integrated photodetector. The LD was directly modulated with a sine-wave AC signal, generated by an arbitrary function generator (AFG). A 10-$\Omega$ sensing resistor was connected in series with the LD to monitor the current change during LD modulation. A bias tee combined the DC and AC LD drive signals. Light coupled into the on-chip SiN waveguide from the bonded LD was split into two paths via the power tap --- the tap path directed a fraction of the light to the integrated photodetector, and the thru path routed light through the VOA to an output fiber-to-chip edge coupler, which was coupled to a single-mode fiber (SMF). The photocurrent of the integrated photodetector was amplified by a transimpedance amplifier (TIA). The PIC output was routed to an off-chip switchable-gain detector through the SMF. The voltage drop across the resistor was measured with an oscilloscope for subsequent current calculations (Chn 1). The amplified photocurrent signal (Chn 2) and output power signal (Chn 3) were simultaneously collected by the oscilloscope. The oscilloscope was internally triggered by the voltage drop across the resistor at Chn 1, ensuring synchronized signal acquisition across the 3 channels.

\subsection{Quantifying dynamic optical power during LD modulation}
\label{calibration}

The dynamic optical power during sinusoidal laser modulation (reported in Section \ref{power monitoring} and Figure \ref{fig8}) was calculated from the measured oscilloscope traces using the following analysis and procedure.

In the DC LD driving case, the relationship between the output optical power from the PIC incident on the off-chip switchable-gain detector ($P_0$) and the measured oscilloscope voltage magnitude ($V_0$) can be described by Equation \ref{eq1}:

\begin{equation}
\label{eq1}
    V_0=G R P_0+F,
\end{equation}
where $G$ and $R$ are the gain and responsivity of the off-chip detector, and $F$ is the DC voltage offset of the oscilloscope trace with no optical power at the detector. Equation \ref{eq1} can be further simplified as: 
\begin{equation}
\label{eq2}
    V_0=a P_0+b,
\end{equation}
where $P_0$ and $V_0$ follow a linear relationship, $a$ and $b$ being the slope and intercept of the linear function.

During sinusoidal modulation of the LD drive current, the measured time-dependent oscilloscope voltage magnitude, $V(t)$, can be further described by Equation \ref{eq3} (neglecting phase delay between $V$ and $P$):
\begin{equation}
\label{eq3}
    V(t)=V_0+\Updelta V \textrm{sin}(2\pi f t)=a [P_0+|H(f)| \Updelta P \textrm{sin}(2\pi f t)]+b,
\end{equation}
where $\Updelta V$ and $\Updelta P$ are the amplitude of the voltage signal and the corresponding off-chip optical power modulation. $|H(f)|$ is the attenuation factor determined by the 3-dB optoelectronic bandwidth, $BW$, of the detector, and $f$ is the sinusoidal modulation frequency. The dependence of the attenuation factor (represented by the frequency response) on the detector's bandwidth is approximated in Equation \ref{eq4} as follows:
\begin{equation}
\label{eq4} \lvert{H(f)}\rvert=\frac{1}{\sqrt{1+(f/BW)^2}}.
\end{equation}

Comparing Equations \ref{eq2} and \ref{eq3}, the amplitude of the dynamic off-chip optical power during modulation, $\Updelta P$, can be derived from Equation \ref{eq5}:
\begin{equation}
\label{eq5} 
\Updelta P=\frac{\Updelta V}{a \lvert{H(f)}\rvert},
\end{equation}
and the maximum and minimum dynamic off-chip optical power during LD modulation can be calculated as $P_{max}=P_0+\Updelta P$ and $P_{min}=P_0-\Updelta P$, where $P_0$ represents the mean of the dynamic optical power. 

To determine the parameters $a$ and $b$, $V_0$ (at oscilloscope Chn3, Figure \ref{figS7}) was recorded for a series of DC LD drive currents, while $P_0$ was measured at each of these drive currents using a power meter. Following Equation \ref{eq2}, a linear fit was performed to obtain $a$ and $b$. $|H(f)|$ at $f = 1$ MHz was then calculated from the specified optoelectronic bandwidth of the detector ($BW = 1.6$ MHz in this work) using Equation \ref{eq4}. Finally, with $a$, $|H(1 MHz)|$, and $b$, Equation \ref{eq3} was applied to compute $P_{max}$, $P_0$, and $P_{min}$ from $V(t)$ during 1 MHz sinusoidal modulation of the LD drive current.  

\twocolumn

\bibliographystyle{IEEEtran}



\end{document}